\documentclass[traditabstract, final]{aa}
\usepackage{graphicx}
\usepackage[small]{caption}
\usepackage{epsfig}
\usepackage{natbib}
\usepackage{fancyhdr}
\usepackage{amsmath}
\usepackage{amssymb}
\usepackage{makeidx}
\usepackage{titletoc}
\usepackage[]{subfigure}
\usepackage{lscape}
\usepackage{hyperref}
\usepackage{wasysym}
\usepackage{wrapfig}
\usepackage{longtable}
\usepackage{aas_macros}
\usepackage{txfonts}

\begin{document}

\title{The Star Formation \& Chemical Evolution History of the Fornax Dwarf Spheroidal Galaxy\thanks{Table 2 is only available in electronic form at the CDS via anonymous ftp tocdsarc.u-strasbg.fr (130.79.128.5) or via http://cdsweb.u-strasbg.fr/cgi-bin/qcat?J/A+A/}}
\titlerunning{Star Formation History of the Fornax dSph}

   \author{T.J.L. de Boer\inst{1} \fnmsep \thanks{Visiting astronomer, Cerro Tololo Inter-American Observatory, National Optical Astronomy Observatory, which are operated by the Association of Universities for Research in Astronomy, under contract with the National Science Foundation.}
          \and
          E. Tolstoy\inst{1} 
          \and
          V. Hill\inst{2}
          \and
          A. Saha\inst{3}
          \and
          E.W. Olszewski\inst{4} \fnmsep \footnotemark[2] 
          \and
          M. Mateo\inst{5} 
          \and
          E. Starkenburg\inst{6}
          \and
          G. Battaglia\inst{7} 
          \and
          M.G. Walker\inst{8} 
  }
   \offprints{T.J.L. de Boer}

   \institute{Kapteyn Astronomical Institute, University of Groningen,
              Postbus 800, 9700 AV Groningen, The Netherlands\\
              \email{deboer@astro.rug.nl}
             \and
              Universit\'{e} de Nice Sophia-Antipolis, CNRS, Observatoire de C\^{o}te d$'$Azur, Laboratoire Cassiop\'{e}e, 06304 Nice Cedex 4, France 
             \and
              National Optical Astronomy Observatory\thanks{The National Optical Astronomy Observatory is operated by AURA, Inc., under cooperative agreement with the National Science Foundation.},
              P.O. box 26732, Tucson, AZ 85726, USA
             \and
             Steward Observatory, The University of Arizona, Tucson, AZ 85721, USA
             \and
             Department of Astronomy, University of Michigan, Ann Arbor, MI 48109-1090, USA
             \and
             Department of Physics and Astronomy, University of Victoria, 3800 Finnerty Road, Victoria, BC, V8P 1A1, Canada
             \and
             INAF~$-$~Osservatorio Astronomico di Bologna Via Ranzani 1, I$-$40127, Bologna, Italy
             \and
             Harvard-Smithsonian Center for Astrophysics, 60 Garden Street, Cambridge, MA 02138, USA
             }

   \date{Received ...; accepted ...}

\abstract{We present deep photometry in the B,V and I filters from CTIO/MOSAIC for about 270.000 stars in the Fornax dwarf Spheroidal galaxy, out to a radius of r$_{ell}$$\approx$0.8 degrees. By combining the accurately calibrated photometry with the spectroscopic metallicity distributions of individual Red Giant Branch stars we obtain the detailed star formation and chemical evolution history of Fornax. \\
Fornax is dominated by intermediate age~(1$-$10 Gyr) stellar populations, but also includes ancient~(10$-$14 Gyr), and young~($\le$1 Gyr) stars. We show that Fornax displays a radial age gradient, with younger, more metal-rich populations dominating the central region. This confirms results from previous works. Within an elliptical radius of 0.8 degrees, or 1.9 kpc from the centre, a total mass in stars of 4.3$\times$10$^{7}$ M$_{\odot}$ was formed, from the earliest times until 250 Myr ago. \\
Using the detailed star formation history, age estimates are determined for individual stars on the upper RGB, for which spectroscopic abundances are available, giving an age-metallicity relation of the Fornax dSph from individual stars. This shows that the average metallicity of Fornax went up rapidly from [Fe/H]$\le$$-$2.5 dex to [Fe/H]=$-$1.5 dex between~8$-$12 Gyr ago, after which a more gradual enrichment resulted in a narrow, well-defined sequence which reaches [Fe/H]$\approx$$-$0.8 dex, $\approx$3 Gyr ago. \\
These ages also allow us to measure the build-up of chemical elements as a function of time, and thus determine detailed timescales for the evolution of individual chemical elements. A rapid decrease in [Mg/Fe] is seen for the stars with [Fe/H]$\ge$$-$1.5 dex, with a clear trend in age.}


\maketitle

\section{Introduction}
\label{Fnxintroduction}
Due to their close proximity to the Milky Way~(MW), dwarf spheroidal galaxies~(dSph) can easily be resolved into individual stars, and studied in detail. Once thought to be the simplest systems in the Local Group~(LG), the dSphs have been found to display complicated evolutionary histories~\citep[e.g.,][]{Tolstoy09}. \\
Fornax is a well studied galaxy, discovered by Shapley in 1938, and is one of the most luminous and massive dSphs in the Local Group, second only to the Sagittarius dwarf~\citep{Mateo98}. It has an absolute magnitude of M$_{\mathrm{V}}$=$-$13.3, and a total~(dynamical) mass of $\approx$1.6$\times$10$^{8}$ M$_{\odot}$, from modelling of kinematic data out to a radius of 1.9 kpc~\citep{Walker06, Klimentowski07, Lokas09}. Fornax is located at a distance of~138$\pm$8 kpc~((m-M)$_{\mathrm{V}}$=20.84$\pm$0.04), determined using RR Lyrae stars, in good agreement with other measurements using the infrared tip of the Red Giant Branch~(RGB) method and the Horizontal Branch~(HB) level~\citep{Greco05,Rizzi07,Pietrzynski09}. The estimated reddening toward Fornax, from extinction maps, is E(B$-$V)=0.03~\citep{Schlegel98}. \\
Fornax can easily be resolved into individual stars, but given the substantial size on the sky~(r$_{\mathrm{tidal}}$=71$\pm$4 arcmin or 2.85$\pm$0.16 kpc), wide-field imaging is needed to study the spatial distribution of the resolved stars~\citep[e.g.,][]{Irwin95, Battaglia06}. The first structural study of Fornax was made with photographic plates by~\citet{Hodge61b}, revealing that the ellipticity of Fornax isophotes increases with distance from the centre. Furthermore, an asymmetry was found with a peak density offset from the central position~\citep{deVaucouleurs68, Hodge74}. Unusually for LG dwarf galaxies, several globular clusters~(GC) were also found~\citep{Shapley382, Hodge61a}, around the Fornax dSph. \\
CCD studies were later conducted in the Fornax dSph, down to ever decreasing brightness levels~\citep{Eskridge883, Eskridge884, Demers94, Irwin95, Walcher03}. The presence of old stellar populations~($\ge$10 Gyr) was confirmed by the discovery of several RR Lyrae stars~\citep{Bersier02}, and the presence of an extended HB~\citep{Saviane00}. Photometric studies have also revealed a large number of carbon stars, indicating a substantial intermediate age~(1$-$10 Gyr) population~\citep{Aaronson80, Aaronson85,Azzopardi99}. In the central regions of Fornax, young Main Sequence stars~($\le$1 Gyr) were first identified by~\citet{Buonanno85}, which led to the discovery of a stellar population gradient of the young and intermediate age stars~\citep{Stetson98}. Photometric surveys of Fornax have found stellar over-densities and ``shells", some of which are believed to be the result of a recent encounter with a smaller system~\citep{Coleman04,Coleman052,Olszewski06}.  \\
The Star Formation History~(SFH) of Fornax has been determined using deep Colour-Magnitude Diagrams~(CMDs)~\citep{Stetson98, Buonanno99, Saviane00, Gallart052}. Using wide-field CMDs,~\citet{Stetson98} determined the age of the young MS population to be 100$-$200 Myr, with similar results found by~\citet{Saviane00}. Deep photometry in small fields close to the centre of Fornax indicate that star formation started $\approx$12 Gyr ago, continuing almost to the present day, with a young episode of star formation 1$-$2 Gyr ago~\citep{Buonanno99,Gallart052}. Furthermore, hints were found of separate Sub-Giant Branches, indicating that star formation may have occurred in discrete episodes~\citep{Buonanno99}. \\
A deep photometric survey was made by~\citet{Coleman08}, covering an area of 5.25 deg$^{2}$ centred on Fornax, down to a 50\% completeness limit of B=23.0 and R=23.5. The SFH was determined at different radii from the centre, showing that Fornax experienced a complex evolutionary history with numerous epochs of star formation. A significant population gradient was found with radius, with ancient stars being present at all radii, but more recent star formation concentrated to the centre. The star formation is found to have gradually decreased from the earliest times, until a sudden episode of strong star formation occurred $\approx$4 Gyr ago. \\
Numerous spectroscopic studies have also been performed in Fornax, of individual stars on the upper RGB. Medium resolution \ion{Ca}{ii} triplet spectroscopy has provided [Fe/H] measurements for increasing numbers of stars~[33 stars~\citep{Tolstoy01}, 117 stars~\citep{Pont04}, 870 stars~\citep{Battaglia082}, 675 stars~\citep{Kirby10}]. These studies provide a detailed Metallicity Distribution Function~(MDF) of Fornax, which shows that the dominant population is relatively metal-rich~([Fe/H]$\approx$$-$0.9 dex), but that stars as metal-poor as [Fe/H]$\le$$-$3.7 dex are also present~\citep{Tafelmeyer10}. By analyzing the spatial distribution of stars in different age ranges, it became clear that the ancient stars display a more extended spatial distribution~(with colder kinematics) than the intermediate age stars, and the latter of the younger stars~\citep{Battaglia06}. \\ 
High resolution~(HR) spectroscopic studies of individual RGB stars in Fornax have been carried out within the central 0.3 degrees of Fornax, for 3 stars~\citep{Tolstoy03,Shetrone03}, 9 stars in three globular clusters~\citep{Letarte06} and more recently 81 stars~\citep{Letarte10} using VLT/FLAMES. These studies have revealed the complex abundance pattern for individual stars in Fornax, including alpha, iron-peak and heavy-elements. 
\begin{figure}[!ht]
\centering
\includegraphics[angle=270, width=0.48\textwidth]{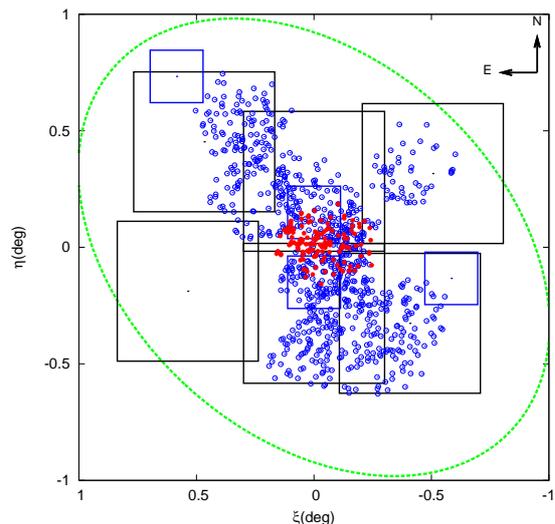}
\caption{Coverage of the photometric and spectroscopic observations across the Fornax dwarf spheroidal galaxy. The big squares denote the coverage of the CTIO 4m/MOSAIC fields, while the smaller squares show the CTIO 0.9m fields. The open~(blue) circles show the stars observed in the VLT/FLAMES \ion{Ca}{ii} triplet survey~\citep{Battaglia082,Starkenburg10}, for which only [Fe/H] is determined. The solid~(red) dots mark the stars observed using medium~\citep{Kirby10} and high resolution spectroscopy~\citep{Letarte10}, for which both [Fe/H] as well as $\alpha$-element abundances are determined. The dashed~(green) ellipse is the tidal radius of Fornax, as given by~\citet{Battaglia06}. \label{Fnxcov}} 
\end{figure}
\\
The observed properties of the Fornax dSph have been modelled several times, in simulations using different techniques, making Fornax one of the best studied dSphs in the Local Group~\citep[e.g.,][]{ Lanfranchi03, Mashchenko08, Revaz09, Kirby11,Revaz11}. The chemo-dynamical Smoothed-Particle Hydrodynamics~(SPH) code used by~\citet{Revaz11} correctly matches the observed metal-rich MDF of Fornax, and shows a narrow [Mg/Fe] distribution, consistent with trends derived from HR spectroscopy. Cosmological simulations have reproduced the stellar parameters of a model galaxy in good agreement with the Fornax dSph, including a globular cluster system~\citep{Mashchenko08}. Furthermore, chemical evolution models of the Fornax dSph have been able to reproduce the observed MDF and alpha-element distribution~\citep[e.g.,][]{Lanfranchi03,Kirby11}. \\
In this work, we present new accurately calibrated, wide-field photometry of the Fornax dSph, going down to the Main Sequence Turn-Off~(MSTO), carried out with MOSAIC on the CTIO 4m/Blanco telescope. Using the accurate photometry, the SFH is determined over a large area of  the Fornax dSph, using CMD synthesis methods~\citep[e.g.,][]{Tosi91,Tolstoy96,Gallart962,Dolphin97,Aparicio97, deBoer2012A}. Spectroscopic metallicities~(from \ion{Ca}{ii} triplet spectroscopy) are directly used in combination with the photometry, to provide additional constraints on the age of the stellar populations, as desribed in~\citet{deBoer2012A}. The detailed SFH is used to determine the probability distribution function for age of individual stars, giving age estimates for stars on the upper RGB. These ages are linked to observed spectroscopic metallicities and abundances, giving an accurate Age-Metallicity Relation~(AMR) and the timescale of chemical enrichment in the Fornax dSph. \\
The paper is structured as follows: in section~\ref{Fnxdata} we present our photometric and spectroscopic observations. Section~\ref{CMDs} presents the CMDs of Fornax and an analysis of the spatial variations of stellar populations within Fornax. In section~\ref{Fnxmethod} we describe the specifics of the method used to obtain the SFH of Fornax. The analysis of the detailed, spatially resolved SFH is given in section~\ref{FnxfullSFH} and the chemical evolution timescales derived from individual RGB stars in section~\ref{Fnxindivages}. Finally, section~\ref{Fnxdiscussion} discusses the results obtained from the SFH and chemical evolution timescales.

\section{Data}
\label{Fnxdata}
\subsection{Photometry}
\label{Fnxphotometry}
Deep optical photometry of the Fornax dSph in the B, V and I filters was obtained using the CTIO 4-m MOSAIC II camera over 9 nights in October 2008 and November 2009. Deep photometry for the outer fields was obtained as part of observing proposal 2008B-0397~(PI M. Mateo) supplemented by deep pointings covering the centre of Fornax as part of observing proposal 2009B-0157~(PI A. Saha). Our observing strategy was to obtain several non-dithered, long~(600s) exposures for each pointing, which were stacked together to obtain the deepest photometry possible. Short~(10s, 90s) exposures were also obtained, in order to sample the bright stars that are saturated in the deep images. \\
In order to ensure accurate photometric calibration of the data set, several fields were observed in Fornax using the 0.9m CTIO telescope, under photometric conditions. Furthermore, observations were also made of Landolt standard fields~\citep{Landolt92, Landolt07} covering a range of different airmass and colour. \\
The positions of the observed fields from the 4m telescope are shown in Figure~\ref{Fnxcov}. The spatial coverage of the B and V filters is complete for r$_{ell}$$\le$0.8 degrees, while the I filters is complete for r$_{ell}$$\le$0.4 degrees. The position, exposure time, airmass and seeing conditions of the Fornax fields are given in Appendix~\ref{Fnxobslist}. \\
The reduction and accurate calibration of this dataset follows the steps described in detail in~\citet{deBoer2011A}. An accurate astrometric solution was determined for each pointing, after which the different exposures of each field were aligned and stacked together to obtain a single, deep image. Photometry was carried out on all images using DoPHOT~\citep{Schechter93}. An accurate photometric calibration solution was determined using the standard star observations, depending on airmass, colour and brightness. All fields were calibrated and subsequently placed on the same photometric scale using the overlapping regions between different fields. Finally, the different fields were combined to obtain a single, carefully calibrated photometric catalog, which is given in table~2.

\subsection{Spectroscopy}
\label{Fnxspectroscopy}
Medium and high resolution spectroscopic observations are available for individual stars on the RGB in Fornax, giving measurements of [Fe/H] and the detailed abundance patterns of individual stars.  \\
Medium resolution~(R$\sim$6500) \ion{Ca}{ii} triplet spectroscopy is available for 870 individual RGB stars in the Fornax dSph, from VLT/FLAMES observations~\citep{Battaglia06, Battaglia082} with the \ion{Ca}{ii} triplet [Fe/H] calibration from~\citet{Starkenburg10}. These observations provide [Fe/H] measurements for stars out to a radius of 1.3 degrees from the centre of the Fornax dSph, and include a range in metallicities from $-$3.0$<$[Fe/H]$<$$-$0.5 dex. Furthermore, medium resolution spectroscopy by~\citet{Kirby10} provides [Fe/H] and $\alpha$-element abundances for 675 stars within the central 0.3 degrees of Fornax, with a metallicity range of $-$2.8$<$[Fe/H]$<$$-$0.1 dex. From this dataset, only those stars with an uncertainty in [Fe/H] and [Mg/Fe] lower than 0.2 dex are used, to select only the brightest stars with a high signal-to-noise. This brings their sample to a total of 72 stars to be considered in the analysis.
\begin{figure}[!ht]
\centering
\includegraphics[angle=0, width=0.45\textwidth]{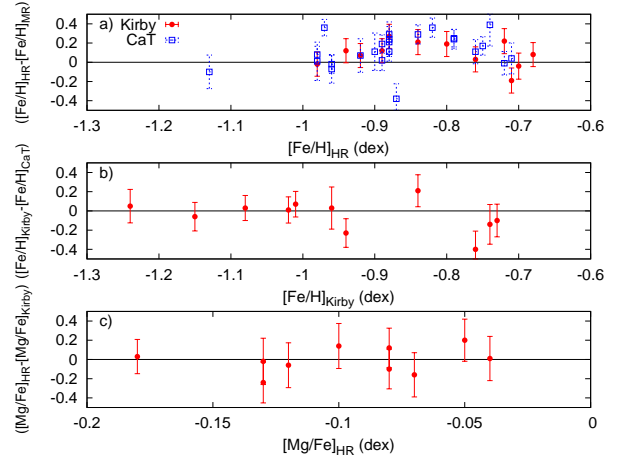}
\caption{Comparison between [Fe/H] and [Mg/Fe] from HR and medium resolution spectroscopy, from~\citet{Letarte10}~(HR),~\citet{Battaglia082}~(CaT) and high S/N stars of~\citet{Kirby10}~(Kirby) using overlapping stars. \textbf{a)} Comparison of the metallicity of HR spectroscopy with high S/N Kirby~(red) and CaT~(blue) stars, as a function of metallicity. \textbf{b)} The difference between medium resolution spectroscopic datasets of Kirby and CaT. \textbf{c)} Comparison between the [Mg/Fe] determined using HR and medium resolution spectroscopy. \label{Fnxcompspec}} 
\end{figure}
\\ 
Furthermore, HR spectroscopy~(from VLT/FLAMES) is also available for r$_{ell}$$\le$0.3 degrees, for 81 individual RGB stars in the Fornax dSph~\citep{Letarte10}, providing [Fe/H] as well as $\alpha$-elements~(O, Mg, Ca, Si, Ti) and r- and s-process elements~(Y, La, Ba, Eu, Nd). For [$\alpha$/Fe] obtained from these observations, we assume [$\alpha$/Fe] =([Mg/Fe] +[Ca/Fe] +[Ti/Fe])/3. The HR spectroscopy covers a range in metallicity from $-$2.6$<$[Fe/H]$<$$-$0.6 dex, for stars on the upper RGB.  \\
The spatial coverage of the spectroscopic observations is shown in Figure~\ref{Fnxcov}, for both the DART HR and \ion{Ca}{ii} triplet spectroscopy as well as the Kirby spectroscopy. \\
To determine if the DART and Kirby spectroscopic catalogs are consistent, we compare [Fe/H] and [Mg/Fe] abundances for overlapping stars. Figure~\ref{Fnxcompspec} shows that the metallicities derived for \ion{Ca}{ii} triplet datasets~\citet{Battaglia082,Starkenburg10} and medium resolution spectroscopy~\citet{Kirby10} are consistent over the metallicity range probed by the overlapping stars. Furthermore, Figure~\ref{Fnxcompspec}c shows that the [Mg/Fe] derived for the highest S/N measurements from~\citet{Kirby10} are consistent with those determined from the HR spectroscopy of~\citet{Letarte10}.
\begin{figure*}[!ht]
\centering
\includegraphics[angle=0, width=0.97\textwidth]{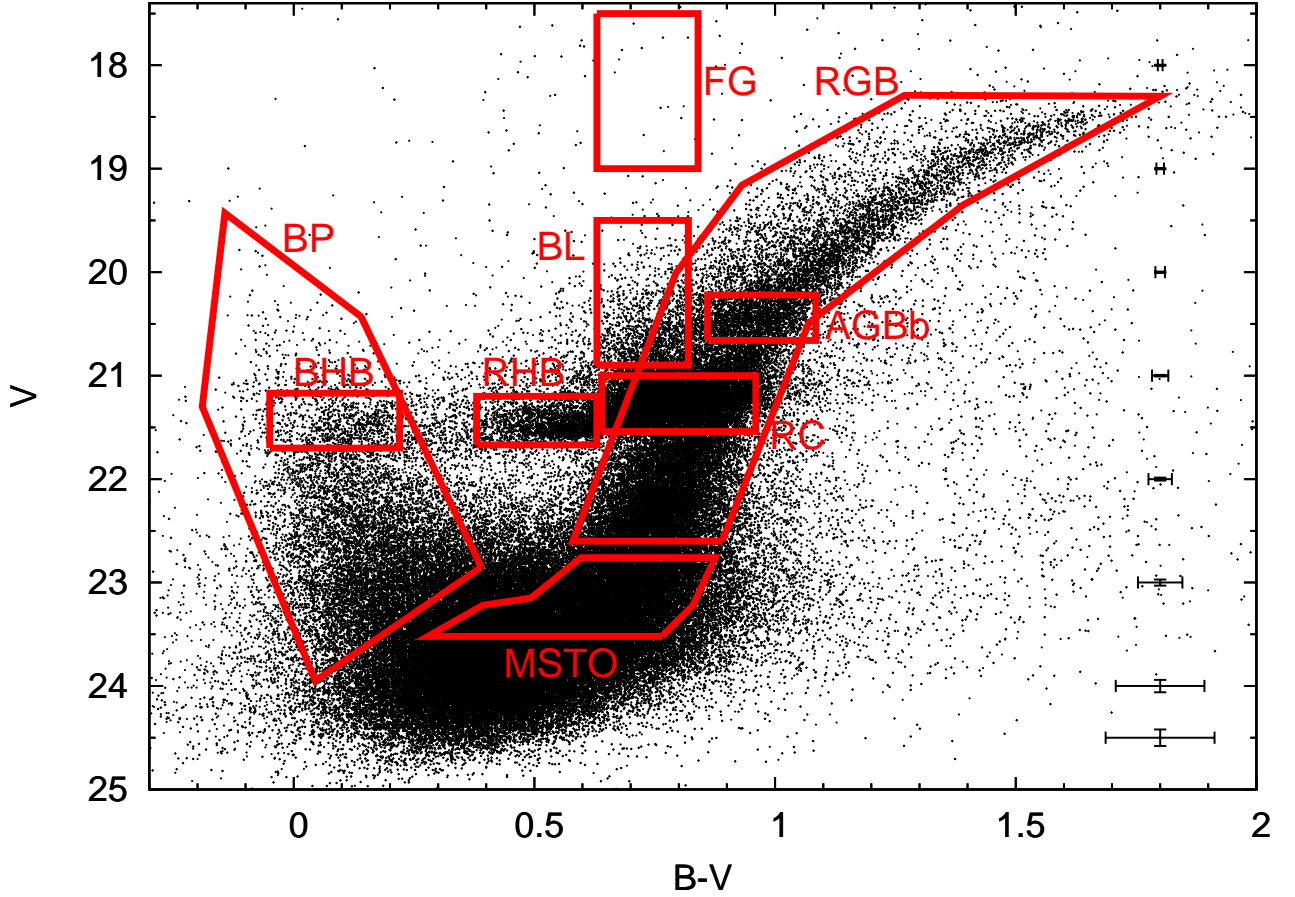}
\caption{Fornax (V, B$-$V) CMD of the central r$_{ell}$$\le$0.4 degrees with boxes identifying the different stellar populations. BHB: Blue Horizontal Branch, RHB: Red Horizontal Branch, RGB: Red Giant Branch, MSTO: Main Sequence Turn-Off, BP: Blue Plume, BL: Blue Loop, RC: Red Clump, AGBb: AGB bump. A box providing an estimate of the Milky Way foreground contamination is also shown~(FG). Error bars represent the photometric errors at each magnitude level.
\label{FnxBVboxes}} 
\end{figure*}

\section{The Colour-Magnitude Diagrams}
\label{CMDs}
The full, calibrated photometry catalog discussed in Section~\ref{Fnxphotometry} contains photometry in the Fornax dSph in the B,V and I filters for radii out  to r$_{ell}$$\approx$1.5 degrees. Due to the spatial coverage of the catalog, the B and V filters are fully sampled for radii r$_{ell}$$\le$0.8 degrees, while the I filter is complete only for r$_{ell}$$\le$0.4 degrees. Figure~\ref{FnxBVboxes} shows a (V, B$-$V) CMD of the inner r$_{ell}$$\le$0.4 degrees of the Fornax dSph, marked with the different features that can be identified in the CMD. \\
The CMD of Fornax displays a wide variety of evolutionary features, consistent with an extended period of star formation. The presence of blue RGB stars and an old MSTO indicates ancient stars~($\ge$10 Gyr), which is consistent with the presence of Blue Horizontal Branch~(BHB) stars. A well populated Red Horizontal Branch~(RHB) and Red Clump~(RC) indicate that the main population of stars in Fornax has intermediate age~(1$-$10 Gyr), further supported by the presence of a strong red RGB population. Clear signs of recent star formation~($\le$2 Gyr) are also found in Figure~\ref{FnxBVboxes}, such as the presence of Blue Loop~(BL) stars and a strong Blue Plume~(BP) population. \\
To determine the photometric completeness of the CMD, a large number of artificial star test simulations have been carried out, similar to the method described in~\citet{deBoer2012A}. A set of artificial stars was generated with parameters within the range 0.25$<$age$<$15 Gyr, $-$2.5$<$[Fe/H]$<$$-$0.30 dex, $-$0.2$<$[$\alpha$/Fe]$<$0.60 dex. Stars were distributed randomly across the six MOSAIC pointings in Fornax, in 3 different filters. To avoid changing the crowding properties in the images no more than 5\% of the total observed stars were ever injected as artificial stars at one time. A total of 400 images in each of the two inner fields, and 200 images in each of the four outer fields were created, containing 5000 artificial stars each. This resulted in 4000 images, containing a total of 8 million artificial stars spread across the full area of the Fornax dSph. 
\begin{figure}[!ht]
\centering
\includegraphics[angle=0, width=0.45\textwidth]{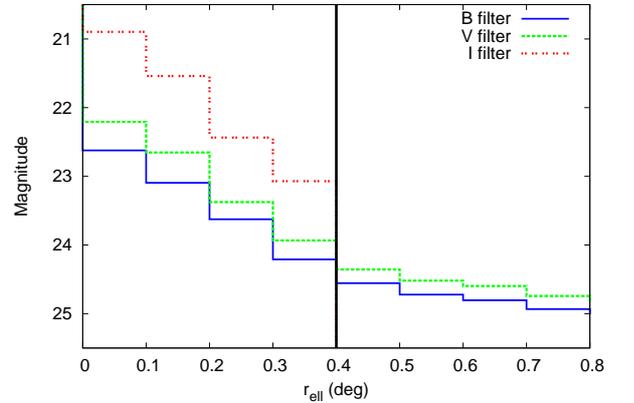}
\caption{The 50\% completeness level in the~B~(solid blue line),~V~(green dashed line) and~I~(red dotted line) filters for increasing elliptical radius~(r$_{ell}$) from the centre of the Fornax dSph. The vertical~(black) line indicates the radius up to which observations in the I filter are completely sampled. \label{Fnxradialcomp}} 
\end{figure}
\\
Figure~\ref{Fnxradialcomp} shows the variation of the 50\% completeness level with distance from the centre of Fornax, in the three different filters. The completeness levels for the B and V filters are comparable at each radius, while the 50\% completeness level in the I filter is systematically brighter due to the increased sky brightness in I. However, the higher intrinsic brightness of stars in the I band compensates for the brighter completeness level. The completeness level in the I filter is shown only for r$_{ell}$$\le$0.4 degrees, due to the lack of I band observations in the outer parts of Fornax. The centre of Fornax is less complete at a fixed magnitude limit than the outer regions, due to the large amount of crowding in the central region, which means fewer stars are unambiguously detected, and placing the 50\% completeness at brighter magnitude levels. 
\begin{figure*}[!hbt]
\centering
\includegraphics[angle=0, width=0.97\textwidth]{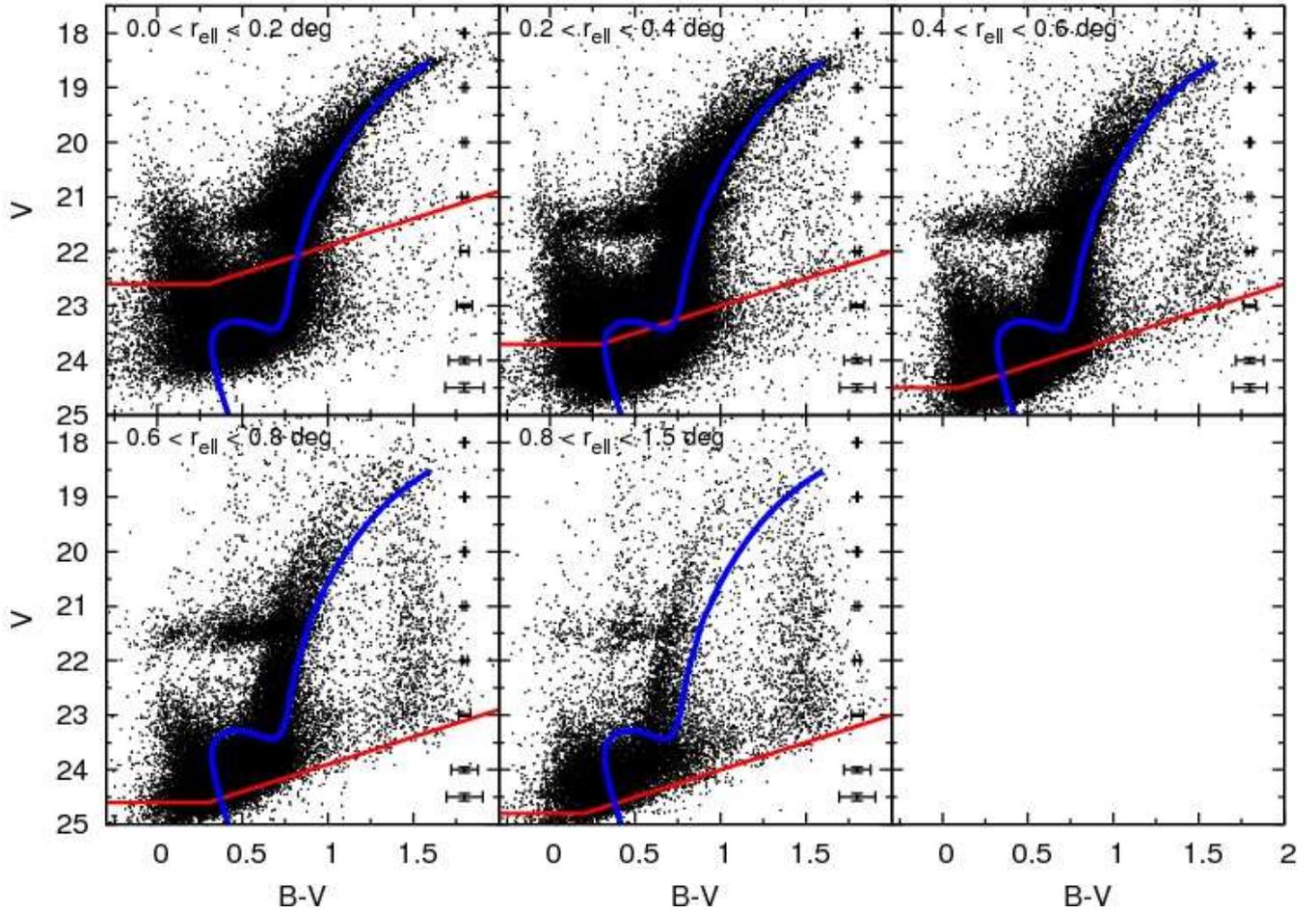}
\caption{(V, B$-$V) Colour-Magnitude Diagrams of the Fornax dwarf spheroidal, shown for increasing elliptical radius. Error bars showing the average photometric error and a 50$\%$ completeness line~(red) are also included. The solid blue line shows a reference isochrone corresponding to the RGB of the dominant population~([Fe/H]=-1.00 dex, [$\alpha$/Fe]=0.00 dex, Age=4 Gyr). \label{FnxBVrad}} 
\end{figure*}
\begin{figure*}[!ht]
\centering
\includegraphics[angle=0, width=0.97\textwidth]{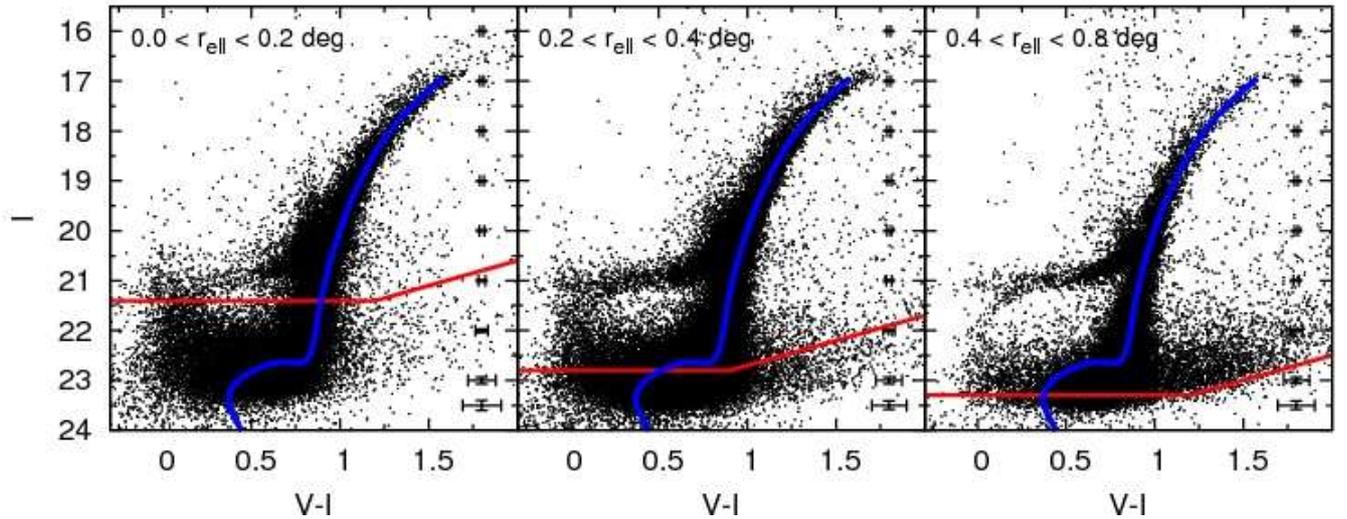}
\caption{(I, V$-$I) Colour-Magnitude Diagrams of the Fornax dSph for increasing elliptical radius. The error bars show the average photometric error at each magnitude level, and the red line indicates the 50$\%$ completeness level. The solid blue line shows a reference isochrone corresponding to the dominant RGB population in Fornax. \label{FnxVIrad}} 
\end{figure*}

\subsection{Spatial variations}
\label{Fnxspatialvariations}
\begin{figure}[!ht]
\centering
\includegraphics[angle=0, width=0.47\textwidth]{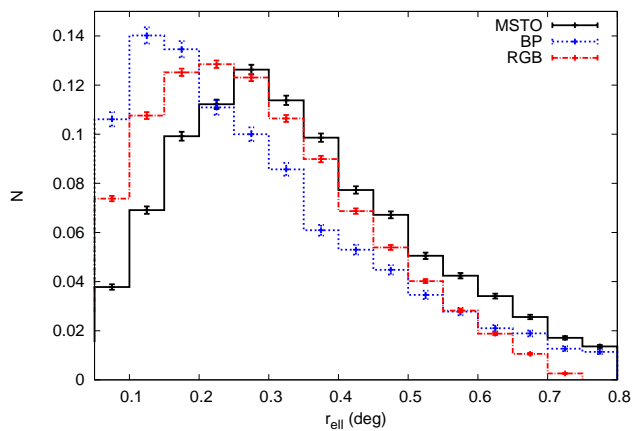}
\caption{Normalised radial distribution of the BP, RGB and MSTO populations~(shown in Figure~\ref{FnxBVboxes}) of the Fornax dSph. \label{radfraction}} 
\end{figure}

Using the full catalog we can construct CMDs at different radii from the centre of the Fornax dSph, which can be used to study the radial behaviour of different evolutionary features. Figures~\ref{FnxBVrad} and~\ref{FnxVIrad} show carefully calibrated~(V, B$-$V) and~(I, V$-$I) CMDs for the Fornax dSph at different elliptical radii~(r$_{ell}$). Error bars show the photometric errors at each magnitude level, and the~(red) line indicates the 50\% completeness level at each elliptical radius. A reference isochrone tracing the dominant RGB of Fornax is also shown, in blue. No de-reddening has been applied to the CMDs in Figures~\ref{FnxBVrad} and~\ref{FnxVIrad}. Instead, models and isochrones used in the CMD analysis are reddened using the extinction coefficient adopted for Fornax~\citep{Schlegel98}. \\
Figures~\ref{FnxBVrad} and~\ref{FnxVIrad} show that the stellar populations in Fornax change with radius from the centre. Ancient stars~($\ge$10 Gyr) are visible at all radii, indicated by the presence of a blue RGB and old MSTO stars. Additionally, the outskirts of Fornax show a well populated BHB population, which is heavily contaminated in the inner parts by the presence of young stars. \\ 
The red RGB and RC populations are dominant in the centre of Fornax, but disappear from the CMDs with increasing elliptical radius. This is a clear sign that the intermediate age stars are more centrally concentrated than the old stars. \\
The bright BP stars are confined to the innermost radial bin in Figures~\ref{FnxBVrad} and~\ref{FnxVIrad}. The outskirts of Fornax do show a population of blue stars extending from the old MSTO, but these stars display a different slope than the BP stars observed in the central region. This feature could be due to the presence of young stars, but is likely caused by the presence of Blue Straggler Stars~(BSS)~\citep{Mapelli09}. The bright BP populations disappear rapidly at larger distances from the centre, which indicates a radially changing age and/or metallicity of stars in Fornax. \\
This is also seen in the radial distribution of the BP, RGB and MSTO populations, as given in Figure~\ref{radfraction}. The young~(BP) population is most centrally concentrated, while the stars on the RGB~(which traces both the intermediate and old populations) are more extended. Finally, the old MSTO stars show an extended radial distribution, with stars present at all radii. To study the stellar populations of Fornax, and their trends, in more detail we analyse the different CMD features and search for variations with elliptical radius. 

\subsubsection{The Red Giant Branch}
\label{FnxRGB}
\begin{figure*}[!ht]
\centering
\includegraphics[angle=0, width=0.95\textwidth]{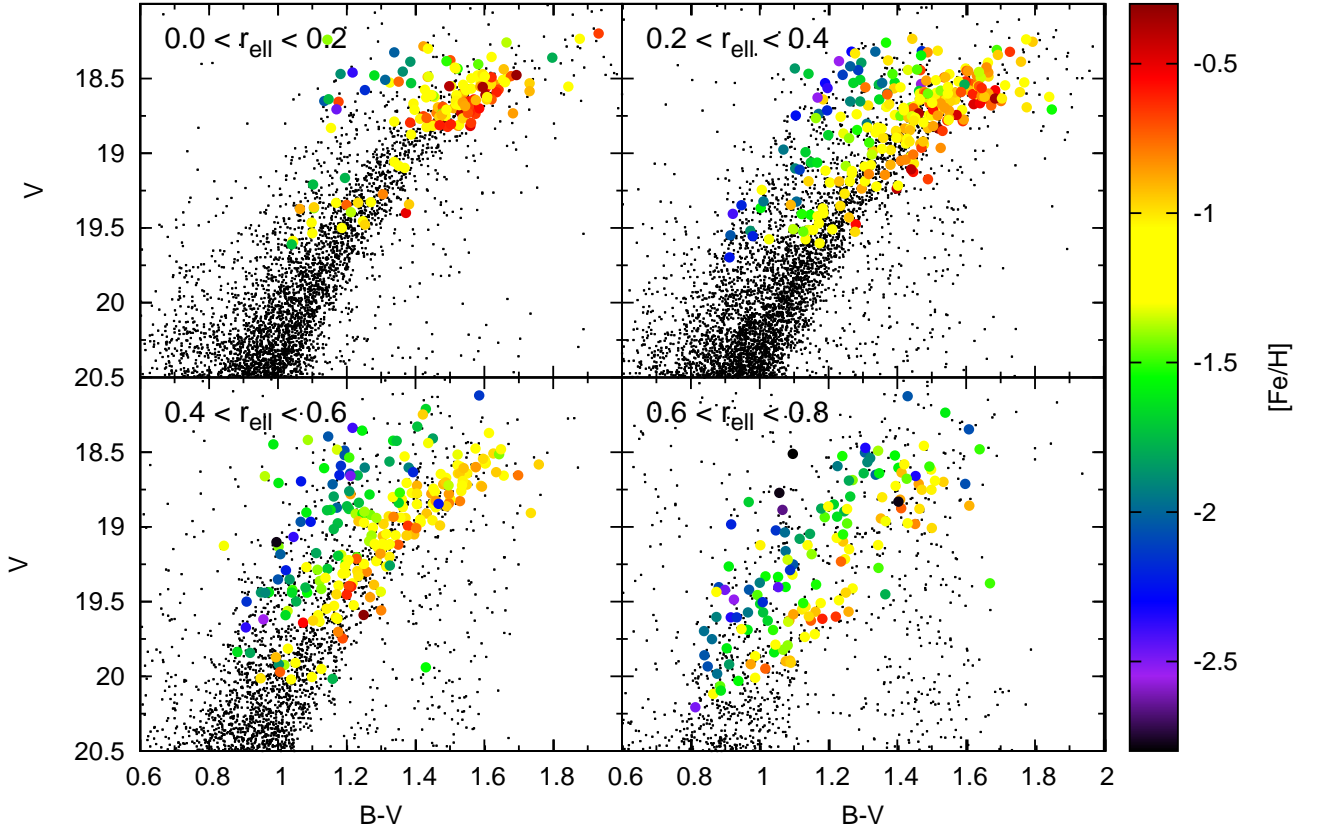}
\caption{The RGB region of the Fornax dSph at different elliptical radii, r$_{ell}$. The~(black) dots show the photometry, while the larger~(coloured) circles indicate the [Fe/H] measurements of individual RGB stars from~\citet{Starkenburg10}. \label{FnxRGBCaT}} 
\end{figure*}

The RGB of Fornax displays a wide spread in colour, as seen in Figures~\ref{FnxBVrad} and~\ref{FnxVIrad}. This is indicative of a spread in metallicity and/or age, as first found by~\citet{Stetson98}. To investigate the properties of RGB stars in detail we use the medium resolution \ion{Ca}{ii} triplet spectroscopy of individual RGB stars from~\citet{Starkenburg10}~(see Section~\ref{Fnxspectroscopy}). Figure~\ref{FnxRGBCaT} shows the RGB of Fornax overlaid with the [Fe/H] metallicities of individual RGB stars where \ion{Ca}{ii} spectroscopy is available. Due to the higher crowding in the central region a brighter magnitude limit was adopted for the spectroscopic observations in the innermost radial bin in Figure~\ref{FnxRGBCaT}.
\\
The inner regions of Fornax show stars with metallicities ranging from $-$2.8$\le$[Fe/H] $\le$$-$0.2 dex. The strong red RGB component seen in Figures~\ref{FnxBVrad} and~\ref{FnxVIrad} shows a metallicity of [Fe/H]$\approx$$-$1.0 dex, and is well traced by an isochrone with an age of 4 Gyr. For increasing radius, Figure~\ref{FnxRGBCaT} shows that the most metal-rich populations~([Fe/H]$\ge$$-$0.7 dex) rapidly disappear. The dominant RGB population also diminishes with radius from the centre, albeit at a slower pace. Conversely, the metal-poor, old populations~([Fe/H]$\le$$-$1.7 dex) remain present at all radii. This observed effect with distance from the centre indicates the presence of a metallicity gradient with radius in Fornax. 

\subsubsection{The Main Sequence Turn-Offs}
\label{FnxMSTO}
\begin{figure*}[!htb]
\centering
\includegraphics[angle=0, width=0.95\textwidth]{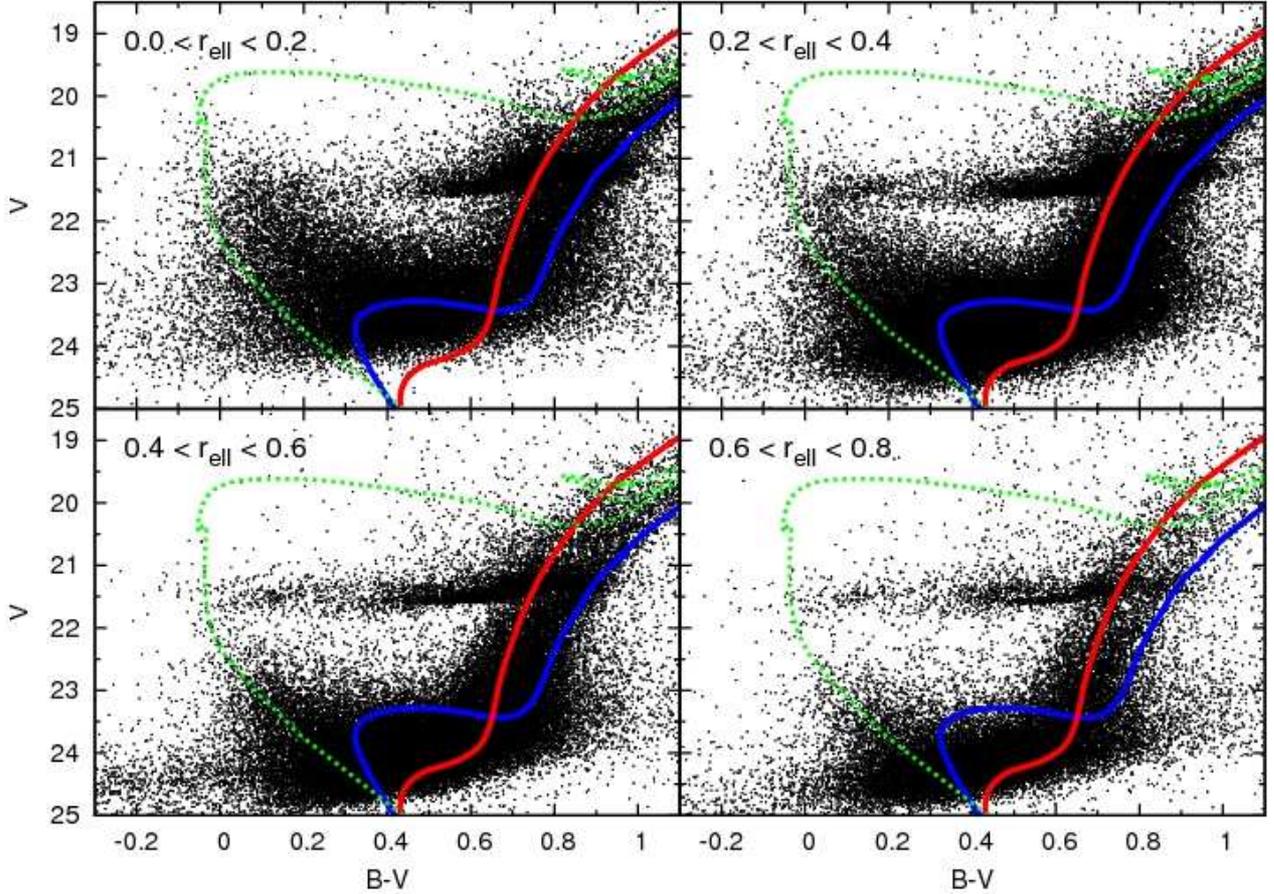}
\caption{The distribution of stars in the MSTO region of the Fornax dSph, for different elliptical radii. Three isochrones are overlaid, representative of the metal-rich~([Fe/H]=$-$1.00 dex, [$\alpha$/Fe]=0.00 dex, age=4 Gyr, red) and metal-poor populations~([Fe/H]=$-$2.45 dex, [$\alpha$/Fe]=0.40 dex, age=14 Gyr, blue)  on the RGB, as well as the young population of the Fornax dSph~([Fe/H]=$-$0.30 dex, [$\alpha$/Fe]=$-$0.20 dex, age=0.3 Gyr, green dotted). \label{FnxradMS}} 
\end{figure*}

The MSTO region of the Fornax dSph~(see Figure~\ref{FnxBVrad}) shows the presence of stellar populations with a wide range of ages, from very young~(BP) to very old~(red MSTO). Previous studies of this region have found clear evidence of a population gradient with radius in the Fornax dSph~\citep{Stetson98, Buonanno99, Gallart052}. Here, we present deep homogeneous photometry of the MSTO region covering a large fraction of the galaxy out to r$_{ell}$=0.8 degrees. \\
Figure~\ref{FnxradMS} shows the MSTO region of the Fornax dSph at different radii from the centre. Reference isochrones have been overlaid on the CMDs, representative of the metallicity of the old, intermediate and young populations present in Fornax~\citep{Battaglia082, Starkenburg10, Letarte10}. The ages of the isochrones have been selected to fit the colours of RGB stars of the corresponding metallicity or the observed distribution of young MS stars. Figure~\ref{FnxradMS} clearly shows that the youngest populations are confined to the centre of Fornax, while the older populations are more extended. The youngest BP populations disappear rapidly as we go outward from the Fornax centre. \\
For increasing elliptical radius the dominant intermediate population also diminishes, as becomes clear when comparing the area underneath the~(red) metal-rich reference isochrone at different distance from the centre. Conversely, the oldest populations traced by the~(blue) metal-poor isochrone remain present at each elliptical radius. A random selection of equal numbers of stars in each panel shows that this effect is not due to the decreased total number of stars at different elliptical radii. These trends are consistent with a radial gradient in both metallicity and age in the Fornax dSph, spanning a wide range in age.

\section{SFH Method}
\label{Fnxmethod}
To determine the SFH of the Fornax dSph, we compare the observed CMDs with a grid of synthetic CMDs using Hess diagrams~(plots of the density of observed stars) of stars in the CMD. We use Talos, which is based on the synthetic CMD method~\citep[e.g.,][]{Tosi91,Tolstoy96,Gallart962,Dolphin97,Aparicio97}. Uniquely, Talos simultaneously takes into account observations in all available photometric filters as well as the spectroscopic MDF, to obtain a well constrained SFH, as discussed in detail in~\citet{deBoer2012A}. \\
First, a set of synthetic ideal CMDs and MDFs is generated using the Dartmouth Stellar Evolution Database~\citep{DartmouthI}, for population bins covering the range in age, metallicity and $\alpha$-element abundances. To simulate observational conditions in the synthetic CMDs used to determine the Fornax SFH, we make use of the artificial star test simulations described in~\ref{CMDs}. This approach is the only way to take into account the complex effects that go into the simulation of observational biases, such as colour-dependence of the completeness level and the asymmetry of the photometric errors of stars at faint magnitudes~\citep[e.g.][]{Gallart961}. The results from the artificial star tests are used to include observational errors in synthetic CMDs in a statistical manner, so they can be compared directly to the observed CMDs, to obtain the best matching SFH.\\
Using the synthetic CMDs, model MDFs are generated by extracting only the stars corresponding to the same magnitude range as the spectroscopic sample. The stars are binned in metallicity and convolved with a Gaussian profile to simulate observational errors.\\
The difference between model and observed CMD and MDF is minimized according to a Poisson equivalent of $\chi^{2}$~\citep{Dolphin02}. The uncertainties on the SFH are obtained by determining the SFH using a series of different CMD and parameter griddings, as described by~\citet{Aparicio09}. The average of all solutions is adopted as the final SFH, with the standard deviation as error bars. In this way, the best matching star formation rate~(SFR) is obtained for each population bin (with age and metallicity) considered in the SFH. These values are then projected onto one axis to obtain the SFR as a function of age~(Star Formation History) or metallicity~(Chemical Evolution History). \\
Talos has been extensively tested, and has been shown to accurately reproduce the properties of both real and synthetic test data~\citep{deBoer2012A}. Before applying Talos to our photometric and spectroscopic data sets of the Fornax dSph, we need to take into account the specifics of the SFH determination of this galaxy. 

\subsection{Parameter space}
The limits of populations adopted in the SFH are determined by considering all available information from the current data sets~(see Section~\ref{Fnxdata}) as well as the literature. 
\begin{figure}[!ht]
\centering
\includegraphics[angle=0, width=0.45\textwidth]{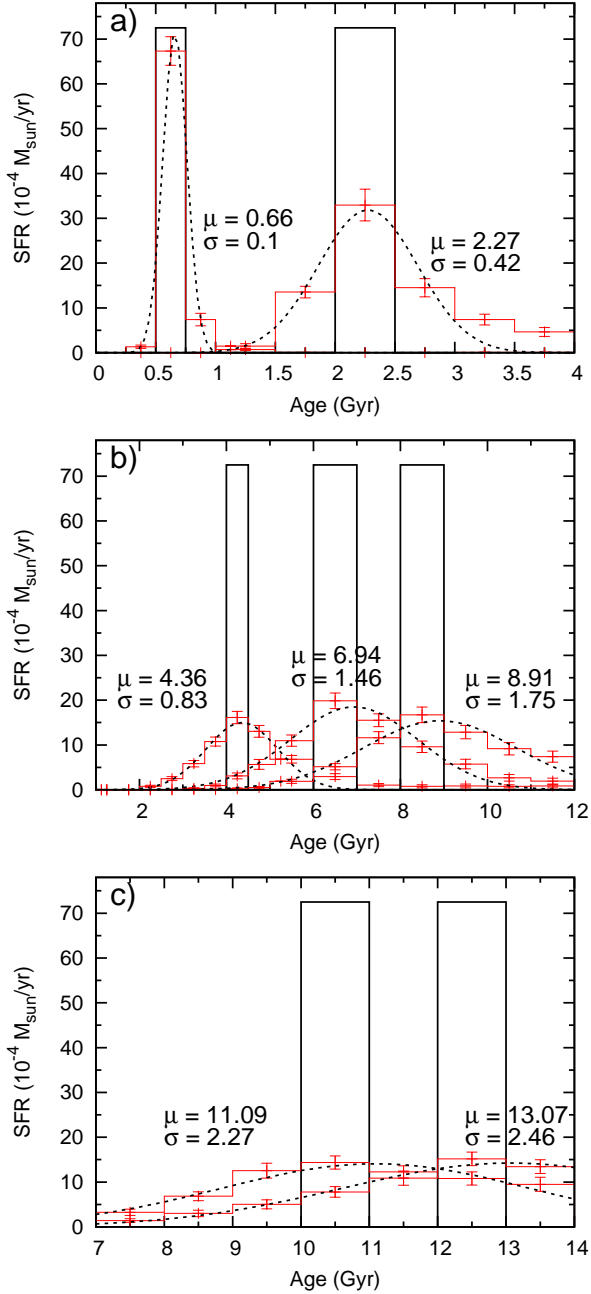}
\caption{Input and recovered SFHs of a series of synthetic short~(10 Myr) episodes of star formation at young~(a), intermediate~(b) and old~(c) ages. The black, solid histogram shows the input SFH, given the adopted binning of the solution. The red histograms show the recovered SFH, as well as the fit of a Gaussian distribution as the dashed line. The mean~($\mu$) and variance~($\sigma$) of the fitted Gaussian distribution are also listed.
\label{Fnxresolution}} 
\end{figure}
\\
For the limits on metallicity, we consider the spectroscopic MDF from \ion{Ca}{ii} triplet and HR spectroscopy. On the metal-poor end, the [Fe/H] is limited to [Fe/H]=$-$2.5 dex, by the availability of metallicity in the Dartmouth Stellar Evolution Database~\citep{DartmouthI}. However, this is not a problem, since the MDF shows that almost no stars in the Fornax spectroscopic sample have [Fe/H]$\le$$-$2.5 dex~\citep{Starkenburg10}. On the metal-rich end, the spectroscopic MDF shows that no stars with [Fe/H]$\ge$$-$0.5 dex are present on the RGB. However, higher metallicities may be present in young Main Sequence stars, which have no corresponding RGB. From previous derivations of the SFH, no stars have been found more metal-rich than [Fe/H]=$-$0.1 dex~\citep{Coleman08}, which we adopt as the metallicity limit in our SFH. A binsize of 0.2 dex is assumed for [Fe/H], which is similar to the average observed uncertainty on [Fe/H]. \\
In order to take into account the $\alpha$$-$element abundances, we adopt a linear [Fe/H] vs [$\alpha$/Fe] relation, based on the HR spectroscopic abundances~\citep{Letarte10}. A range in [$\alpha$/Fe] is assumed for each bin in [Fe/H], which reproduces the trends in the HR data. \\
The choice of possible ages to adopt in the SFH solution is motivated by previous derivations of the SFH of the Fornax dSph. Previous work has shown that Fornax contains stars covering all ages, from ancient~($\ge$10 Gyr) to very young~($\le$1 Gyr) ages~\citep{Buonanno99, Gallart052,Coleman08}. This is also consistent with the observed CMDs, which show a prominent blue plume, caused by stars with ages less than 1 Gyr. The lower limit in age is set by the adopted isochrone set, which does not contain isochrones with ages$\le$0.25 Gyr. Assuming a maximum age of 14 Gyr, for the age of the Universe, we consider therefore a range of ages between 0.25 and 14 Gyr old. A binsize of 1 Gyr is used for ages $\ge$5 Gyr, 0.5 Gyr for ages between 1$-$5 Gyr and 0.25 Gyr for ages $\le$1 Gyr, to take into account the different age sensitivity of young MSTOs~\citep{Hidalgo11}.

\subsection{Age resolution}
\label{Fnxageresolution}
The age resolution of the SFH solution determined by Talos is important to understand the limitations of our derived SFH of Fornax. The ability to resolve episodes of star formation at different ages depends mainly on the photometric depth of the observed photometry. Furthermore, the method of determining the uncertainties of the solution~(see Section~\ref{Fnxmethod}) results in a smoothing of the SFH, which gives limits to the resolution of the SFH at different ages. The age resolution of the solutions is determined by the ability to recover the SFH of a set of synthetic populations at different input ages~\citep[e.g.,][]{Hidalgo11}. \\
A set of seven synthetic short episodes of star formation~(with a duration of 10 Myr) was generated, with different input ages covering the age range adopted for Fornax~(0.6, 2.2, 4.2, 6.5, 8.5,10.5 and 12.5 Gyr). The metallicity distribution of the synthetic episodes is chosen to match the observed MDF of Fornax. Using artificial star test results, observational conditions were simulated corresponding to the central region of Fornax. Given that the crowding is highest in this part of the galaxy~(and the resulting photometry the least deep), this will result in upper limits to the age resolution. The SFH for the synthetic episodes is recovered by fitting the photometry in three filters~(B,V and I), simultaneously with a synthetic 50\% complete MDF with similar photometric depth as the~\ion{Ca}{ii} triplet spectroscopy. \\
The recovered and input SFH for the synthetic episodes is shown in Figure~\ref{Fnxresolution}, for young~(a), intermediate~(b) and old~(c) star formation episodes. The recovered SFH is fit by a Gaussian distribution, from which we determine the age of the central peak~($\mu$) as well as the variance~$\sigma$, which determines the resolution of the recovered episode. Figure~\ref{Fnxresolution} shows that the central peak is recovered at the correct age for young ages~($\le$7 Gyr), given the binning adopted for the solutions. For older ages the peak of the synthetic episode is recovered at slightly too old ages. \\
The Gaussian fits show that the age resolution of the recovered episodes is significantly better at young ages than at old ages. For the young ages with small age resolution, the star formation is confined mostly to the central bin, while for the old episodes the star formation is spread out over multiple bins. The episodes are recovered with a resolution of~$\approx$0.1 Gyr at an age of 0.6 Gyr,~$\approx$0.4 Gyr at an age of 2.2 Gyr,~$\approx$0.8 Gyr at an age of 4.2 Gyr~$\approx$1.5 Gyr at an age of 6.5 Gyr,~$\approx$1.8 Gyr at an age of 8.5 Gyr,~$\approx$2.3 Gyr at an age of 10.5 Gyr and~$\approx$2.5 Gyr at an age of 12.5 Gyr, which is consistent with values between 17 and 22\% of the adopted age.

\section{The Star Formation History of the Fornax dSph}
\label{FnxfullSFH}
\begin{figure*}[!ht]
\centering
\includegraphics[angle=0, width=0.45\textwidth]{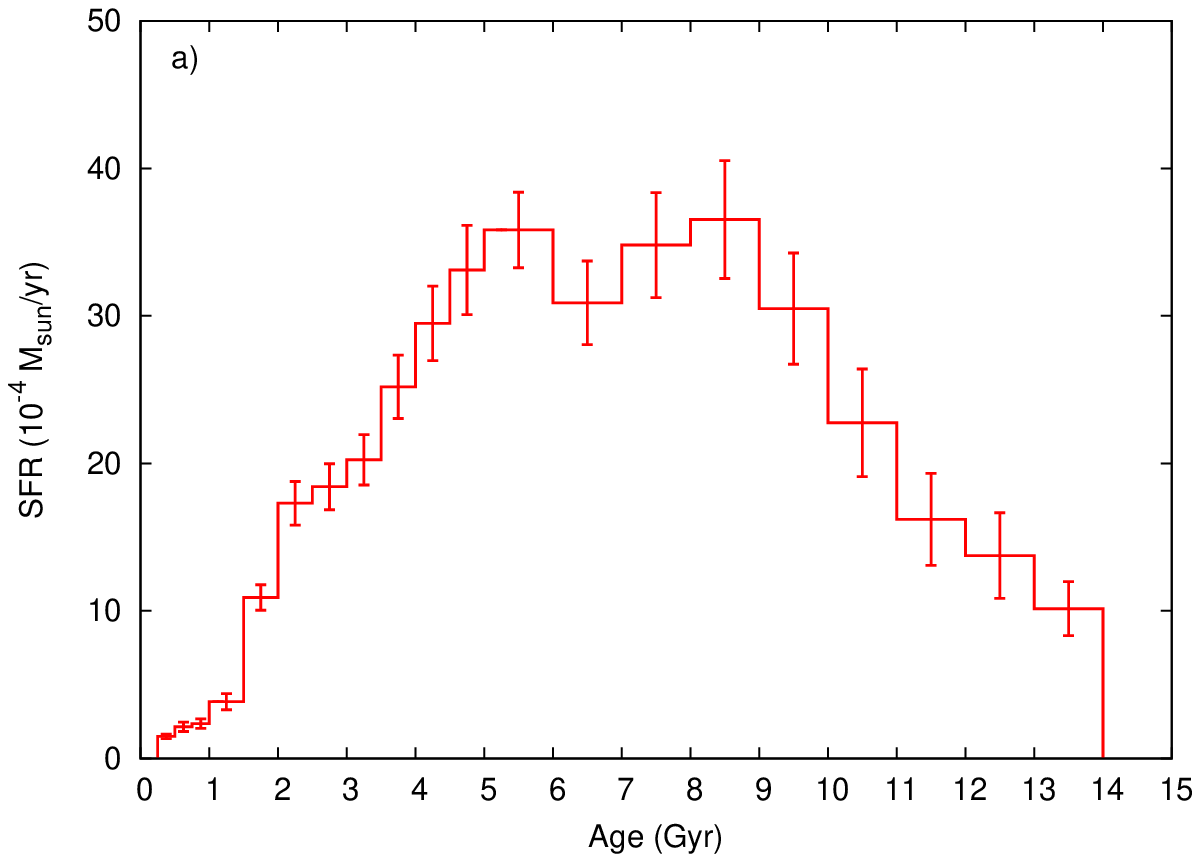}
\includegraphics[angle=0, width=0.45\textwidth]{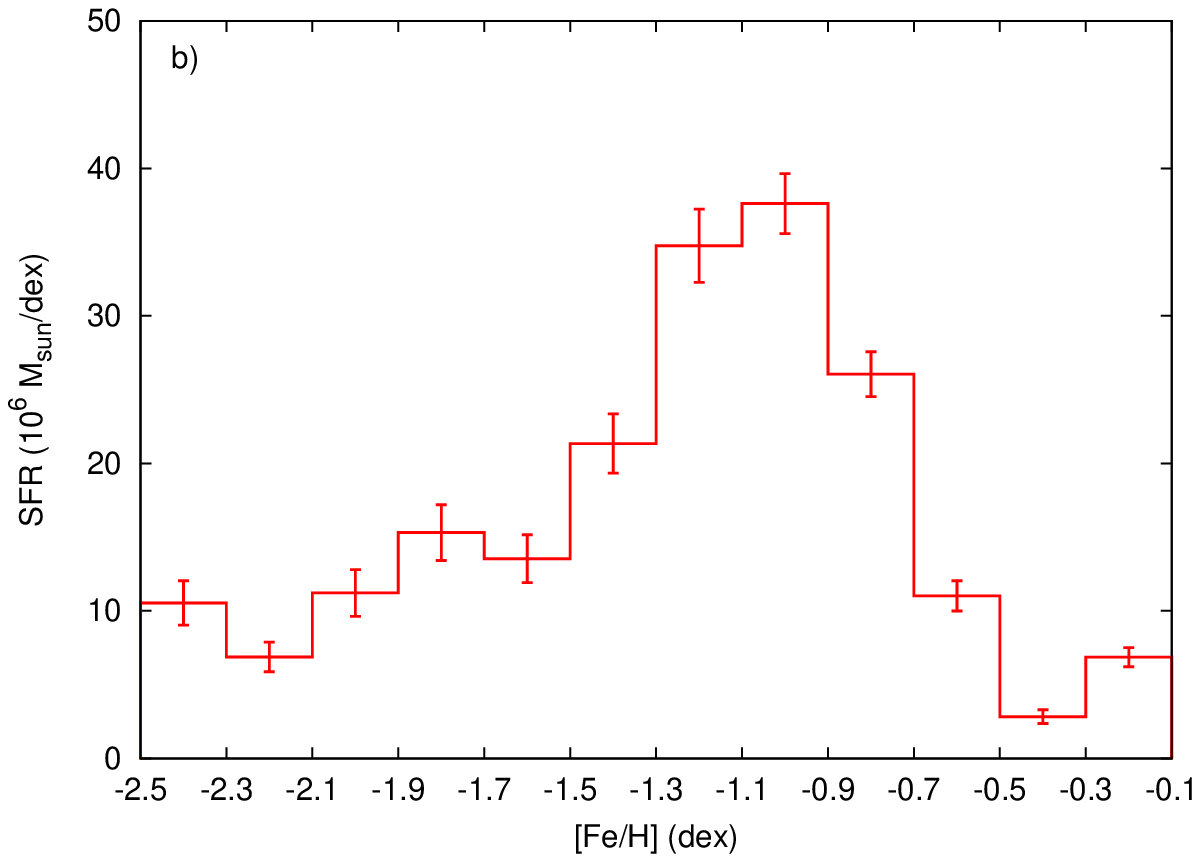}
\caption{The overall~\textbf{a)} Star Formation History and~\textbf{b)} Chemical Evolution History~(CEH) of the Fornax dSph, out to an elliptical radius of r$_{ell}$=0.8 degrees. The SFH and CEH have been determined using all available photometric information, incombination with observed spectroscopic metallicities. \label{FnxoverallSFH}} 
\end{figure*}

\begin{figure}[!ht]
\centering
\includegraphics[angle=0, width=0.45\textwidth]{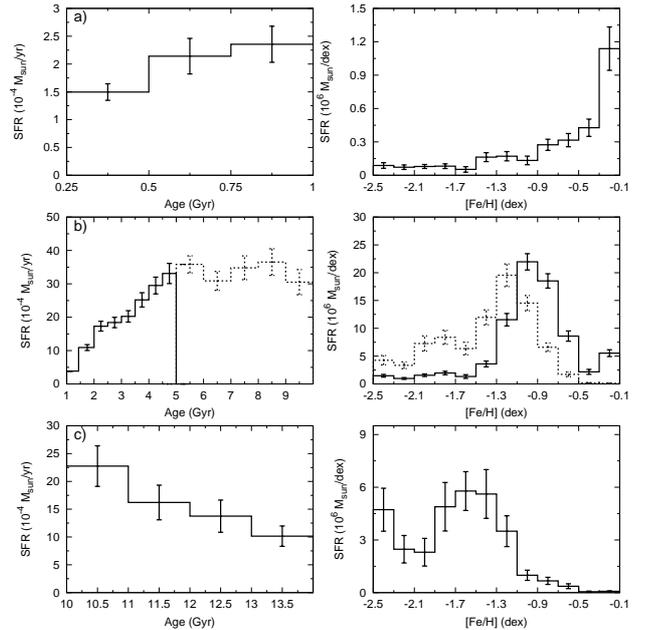}
\caption{The overall SFH of the Fornax dSph, divided into young~(a), intermediate~(b) and old~(c) age ranges. The corresponding CEH is also shown for each age range. For intermediate age stars, younger~(solid) and older~(dashed) stars are separated to show the effect of age on the CEH.  \label{FnxoverallSFHdetail}} 
\end{figure}

\begin{figure*}[!ht]
\centering
\includegraphics[angle=0, width=0.45\textwidth]{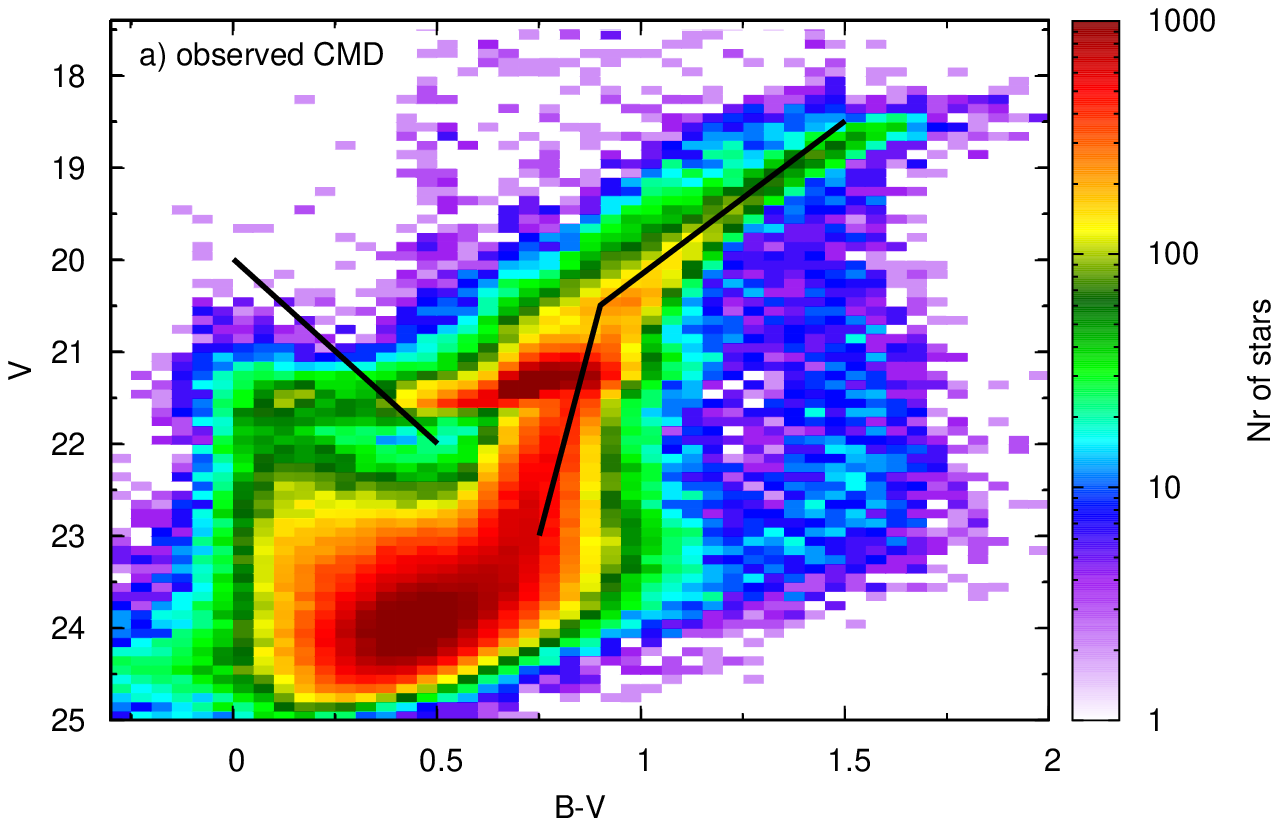}
\includegraphics[angle=0, width=0.45\textwidth]{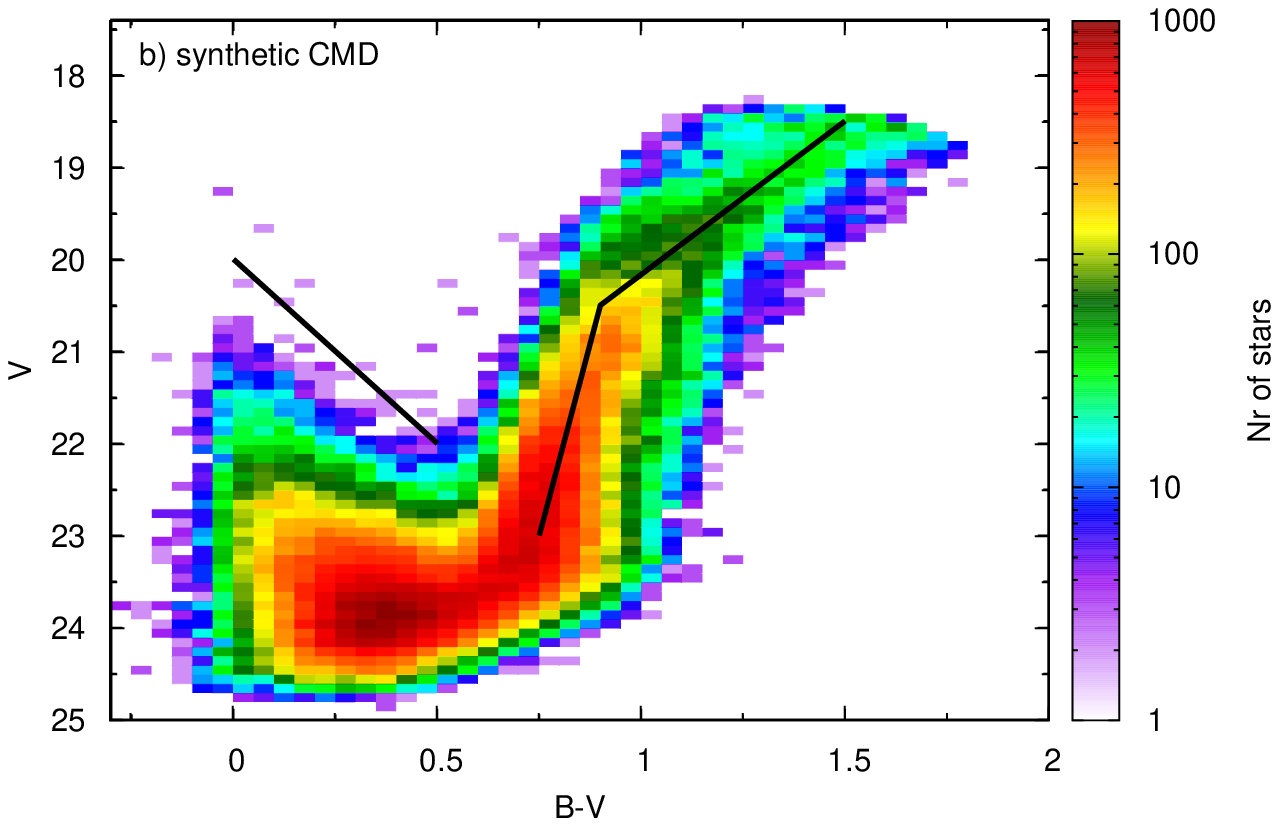}
\caption{The~\textbf{a)} observed and~\textbf{b)} synthetic Hess diagrams for the Fornax dSph, out to r$_{ell}$=0.8 degrees. The colours represent the number of stars in each bin, on a logarithmic scale. To allow a better comparison between both Hess diagrams,~(black) reference lines are also shown. The BHB, RHB and RC are not present in the synthetic Hess diagram, since they are not modelled in the isochrone set used. \label{FnxHesscomparison}} 
\end{figure*}

\begin{figure*}[!ht]
\centering
\includegraphics[angle=0, width=0.47\textwidth]{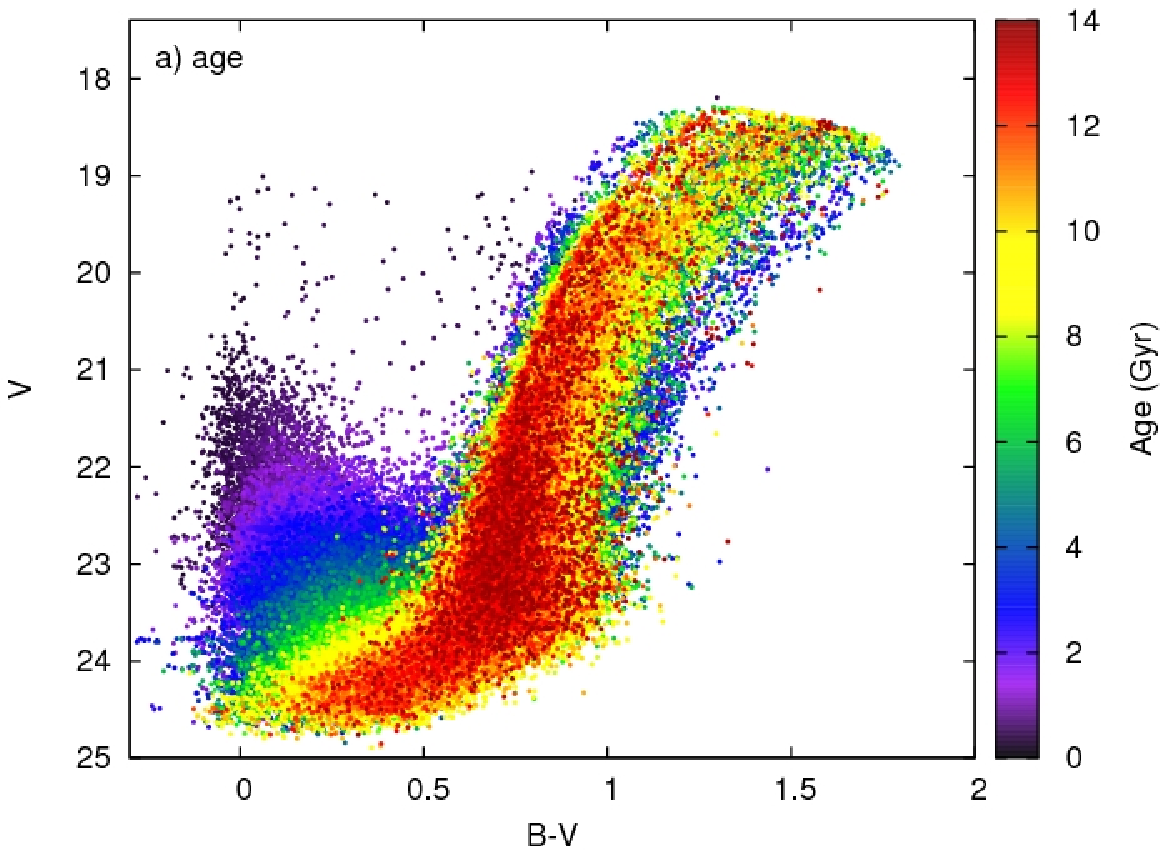}
\includegraphics[angle=0, width=0.47\textwidth]{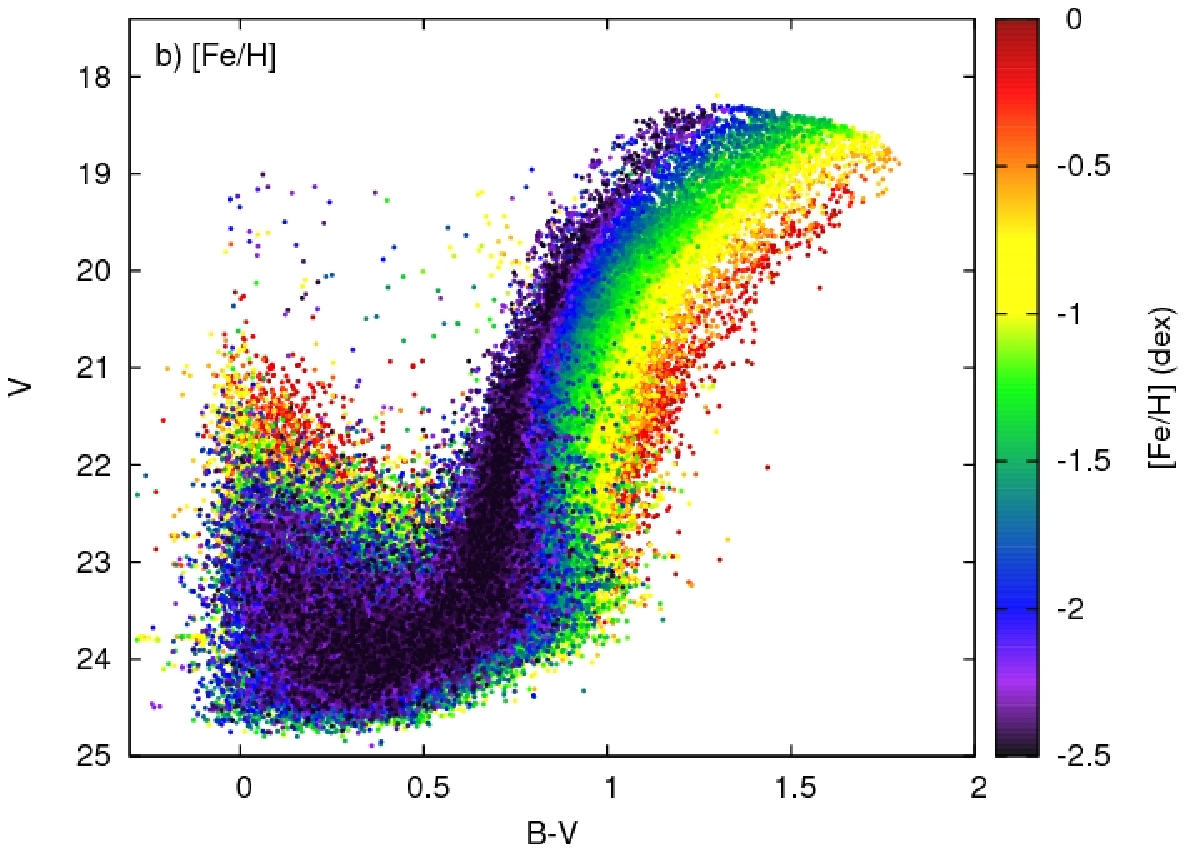}
\caption{The synthetic CMD of the Fornax dSph inferred from the best matching SFH as determined using the Dartmouth isochrone set. The colours represent~\textbf{a)} age and~\textbf{b)} [Fe/H] for each individual synthetic star. \label{Fnxoverlay}} 
\end{figure*}

Having described our method for determining the SFH of the Fornax dSph in Section~\ref{Fnxmethod}, we now apply Talos to our photometric and spectroscopic data sets of the Fornax dSph~\citep{Battaglia082,Starkenburg10, Letarte10, Kirby10}. \\
The SFH is derived using the photometric CMDs in combination with the spectroscopic MDF, to constrain the final SFH as described in~\citet{deBoer2012A}. Within the central 0.4 degrees, photometric information in three filters~(B,V and I) is combined with the available HR and~\ion{Ca}{ii} triplet spectroscopy. For radii r$_{ell}$$\ge$0.4 degrees, the SFH is determined using just the (V, B$-$V) CMD combined with~\ion{Ca}{ii} triplet spectroscopy. The spectroscopic MDF is used to constrain the SFH solution for $-$2.5$\le$[Fe/H]$\le$$-$0.6 dex. CMD analysis shows that more metal-rich stars are too young to appear on the RGB. Therefore, these populations are only constrained by the photometry. An investigation of the MDF at different photometric depth has shown that no luminosity bias is present in the spectroscopic sample of Fornax stars.\\
Figure~\ref{FnxoverallSFH} presents the final SFH and Chemical Evolution History~(CEH) of the Fornax dSph out to a radius of~r$_{ell}$=0.8~degrees, with error bars indicating the uncertainty on the SFR as a result of different CMD and parameter griddings~(as described in Section~\ref{Fnxmethod}). The SFH and CEH display the rate of star formation at different ages and metallicities over the range of each bin, in units of solar mass per year or dex respectively. The total mass in stars formed in each bin can be determined by multiplying the star formation rates by the range in age or metallicity of the bin. \\
The SFH shows that star formation is clearly present at all ages, from as old as 14 Gyr to as young as 0.25 Gyr. Most of the star formation takes place at intermediate ages, between 1$-$10 Gyr, consistent with earlier determinations of the Fornax SFH~\citep{Gallart052,Coleman08}. The CEH in Figure~\ref{FnxoverallSFH}b shows that the dominant population displays metallicities between $-$1.5$\le$[Fe/H]$\le$$-$0.7 dex, but that significant levels of star formation are also present at [Fe/H]$\approx$$-$2.4 dex and [Fe/H]$\approx$$-$0.2 dex. \\
Figure~\ref{FnxoverallSFHdetail} shows the overall SFH of Fornax, divided into young~(a), intermediate~(b) and old~(c) age ranges, with the corresponding CEH. The youngest star formation in Fornax~($\le$1 Gyr) displays declining SFR for younger ages, with a metal-rich CEH showing a peak at [Fe/H]$\approx$$-$0.2 dex. The ages of these populations are determined with a high accuracy~(see Section~\ref{Fnxageresolution}), indicating that the change in SFH corresponds to a real drop in SFR. \\
The intermediate age star formation shows a double peaked SFH with a declining SFR for ages $<$6 Gyr. The resolution of the SFH at these ages~($<$2 Gyr) is good enough to distinguish between both peaks, indicating the presence of distinct episodes of star formation. Furthermore, the age extent of the younger episode~($<$6 Gyr) is inconsistent with a single short burst of star formation and indicates a smooth change in star formation activity over several Gyrs. The metallicities of the populations with ages between 2$-$5 Gyr~(solid) display a range between $-$1.5$\le$[Fe/H]$\le$$-$0.5 dex with a peak at [Fe/H]$\approx$$-$1.0 dex. The slightly older intermediate age star formation~(5$-$10 Gyr, dashed) is dominated by more metal-poor stars with a distribution between $-$2.1$\le$[Fe/H]$\le$$-$0.7 dex, showing two peaks at [Fe/H]$\approx$$-$1.8 and $-$1.2 dex. \\
Finally, the oldest ages~(10$-$14 Gyr) display a declining SFR with increasing age, and a metal-poor CEH with a similar total range as observed in the Sculptor dSph~\citep{deBoer2012A}. However, the SFR at the oldest ages is not constrained well due to the depth of the observed CMDs, resulting in a smoothing of the underlying actual SFH and removing any observable features. \\  
To check the distribution of stars in the CMD we compare the best matching observed and synthetic CMDs of the Fornax dSph in Figure~\ref{FnxHesscomparison}. The synthetic CMD has been generated using the Dartmouth isochrone set~\citep{DartmouthI}, and therefore does not include a RC or HB. The overall distribution of stars in the observed CMD is well matched in the synthetic CMD. Stars on the BP and MSTO occupy the same range in colour and brightness as in the observed CMD. Furthermore, the colour range and slope of the RGB is a good match to the observations, as shown by comparison to the reference lines. \\
The total number of stars in the synthetic CMD is consistent with the observations to within a few percent. Additionally, the relative fraction of stars occupying the RGB and BP phase is similar to that of the observed stars. This shows that the total mass in stars in the Fornax dSph is well matched, and distributed across the CMD with the right numbers. Therefore, we feel confident that the SFH constitutes an accurate representation of the overall behaviour of the stellar populations in the Fornax dSph. 
\\
Using the best matching SFH we can show the age and metallicity corresponding to different CMD features. Figure~\ref{Fnxoverlay} shows the synthetic (V, B$-$V) CMD of Fornax generated using the Dartmouth isochrones, colour coded with age and metallicity for each individual star. The effect of different stellar populations on the MSTO region of the CMD is clearly seen from Figure~\ref{Fnxoverlay}, with younger Main Sequence stars occupying bluer and brighter positions on the BP. The CMD shows that the majority of stars is of intermediate age, but that lower levels of old star formation are also present. The position of different stellar populations on the RGB shows that the oldest stars occupy the blue side of the RGB, while progressively younger stars are found toward the red side of the RGB. However, small numbers of young~($\le$4 Gyr) stars are also found on the blue side of the RGB, after evolving off the young MS. \\
The effect of metallicity on the positions of stellar populations is shown in Figure~\ref{Fnxoverlay}b. The RGB is sensitive to [Fe/H] rather than age, and shows a clear metallicity pattern across the RGB, which provides a good match to the observed metallicity pattern seen in Figure~\ref{FnxRGBCaT}. Stars on the BP are mostly metal-rich, although metal-poor stars are also present. This could be due to the lack of spectroscopy of BP stars, which means the metallicity of these populations is not constrained from the MDF fitting. 

\subsection{The effect of different isochrone sets}
\begin{figure*}[!ht]
\centering
\includegraphics[angle=0, width=0.45\textwidth]{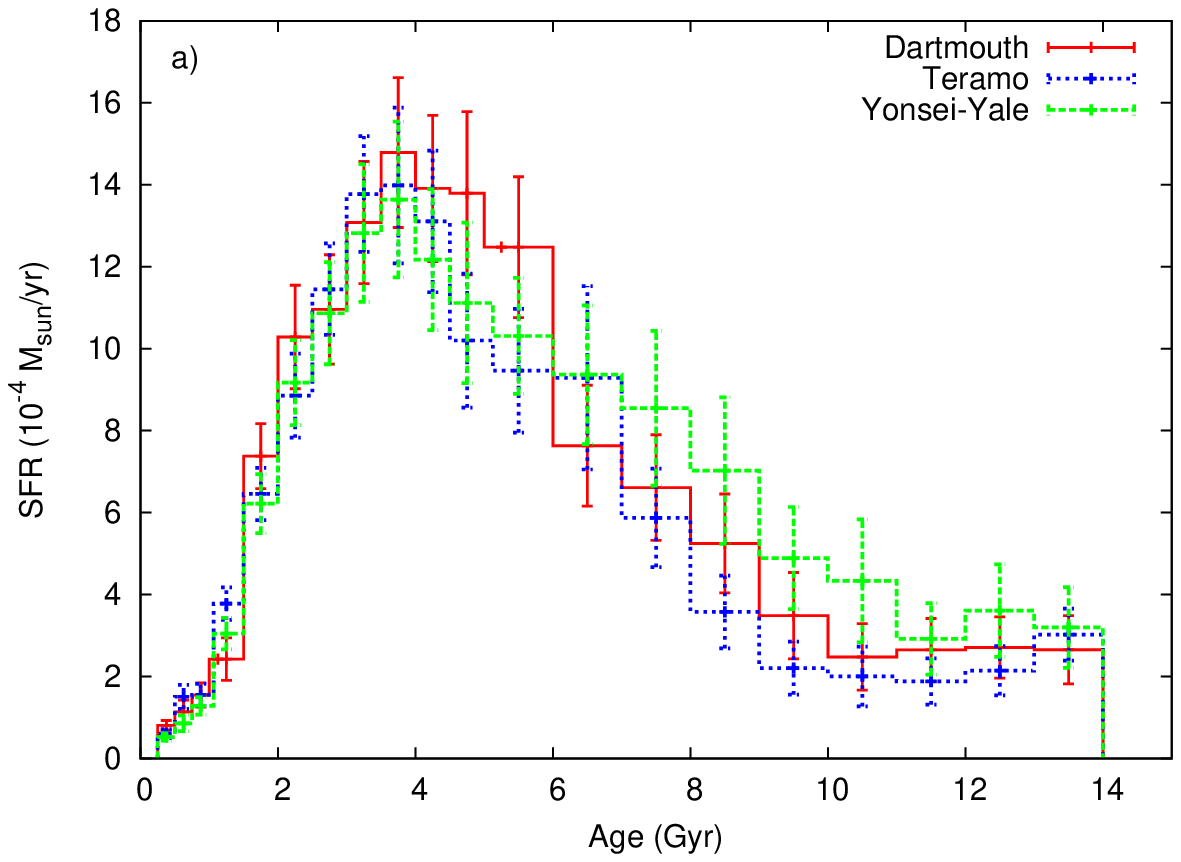}
\includegraphics[angle=0, width=0.45\textwidth]{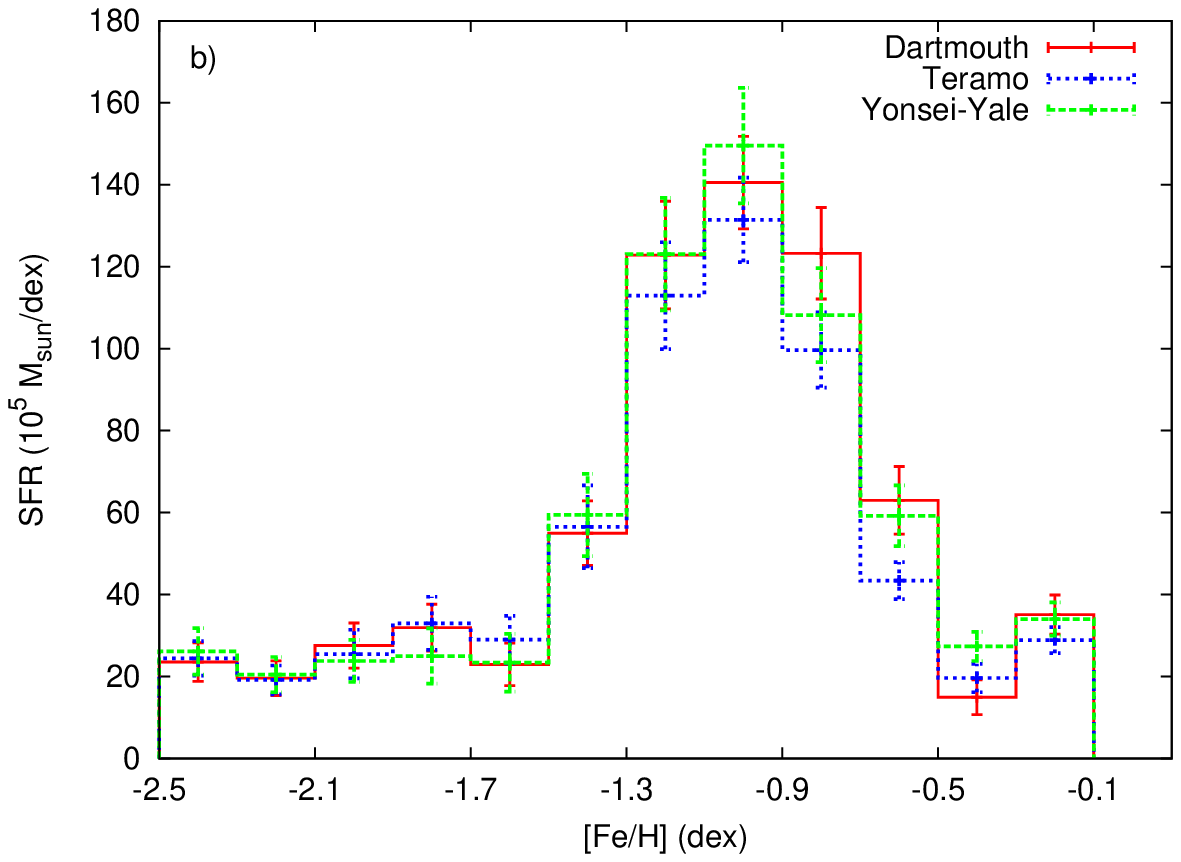}
\caption{Comparison between the~\textbf{a)} SFH and~\textbf{b)} CEH of the centre of Fornax~(r$_{ell}$$\le$0.194 degrees), determined using different isochrone sets. The SFH solution is shown for Darthmouth Stellar Evolution Database isochrones~(red, solid), the BaSTI/Teramo isochrones~(blue, dotted) and the Yonsei-Yale isochrones~(green, dashed). \label{Fnxisodiff}} 
\end{figure*}

Figure~\ref{Fnxisodiff} shows the SFH solution of the central region of Fornax~(r$_{ell}$$\le$0.194 degrees) obtained using three different isochrone sets~\citep{DartmouthI, TeramoI, YonseiYaleI}. The central region was chosen since the range of populations is greatest here, allowing a comparison of determined ages over a large timespan. Figure~\ref{Fnxisodiff} shows that the solution is very similar in all cases, and consistent within the error bars. The overall behavior of the SFH is recovered in all solutions, with a peak at $\approx$4 Gyr, slowly declining SFR toward older ages, and a flat distribution for ages $>$10 Gyr. The similarity between different isochrone sets is likely due to the dominance of the young Main Sequence and MSTO in the solution~(where isochrone agree reasonably well), while the contribution of the RGB~(which is where isochrone sets disagree most) is taken into account solely through the MDF. Therefore, the shape of the RGB is not fit in the solution, but rather the number of stars occupying the upper RGB~(as determined by the isochrone luminosity function). The good agreement of the SFH solutions gives confidence to our final SFH of Fornax. 

\subsection{Spatial variations in the SFH}
\label{Fnxregions}
\begin{figure*}[!ht]
\centering
\includegraphics[angle=0, width=0.97\textwidth]{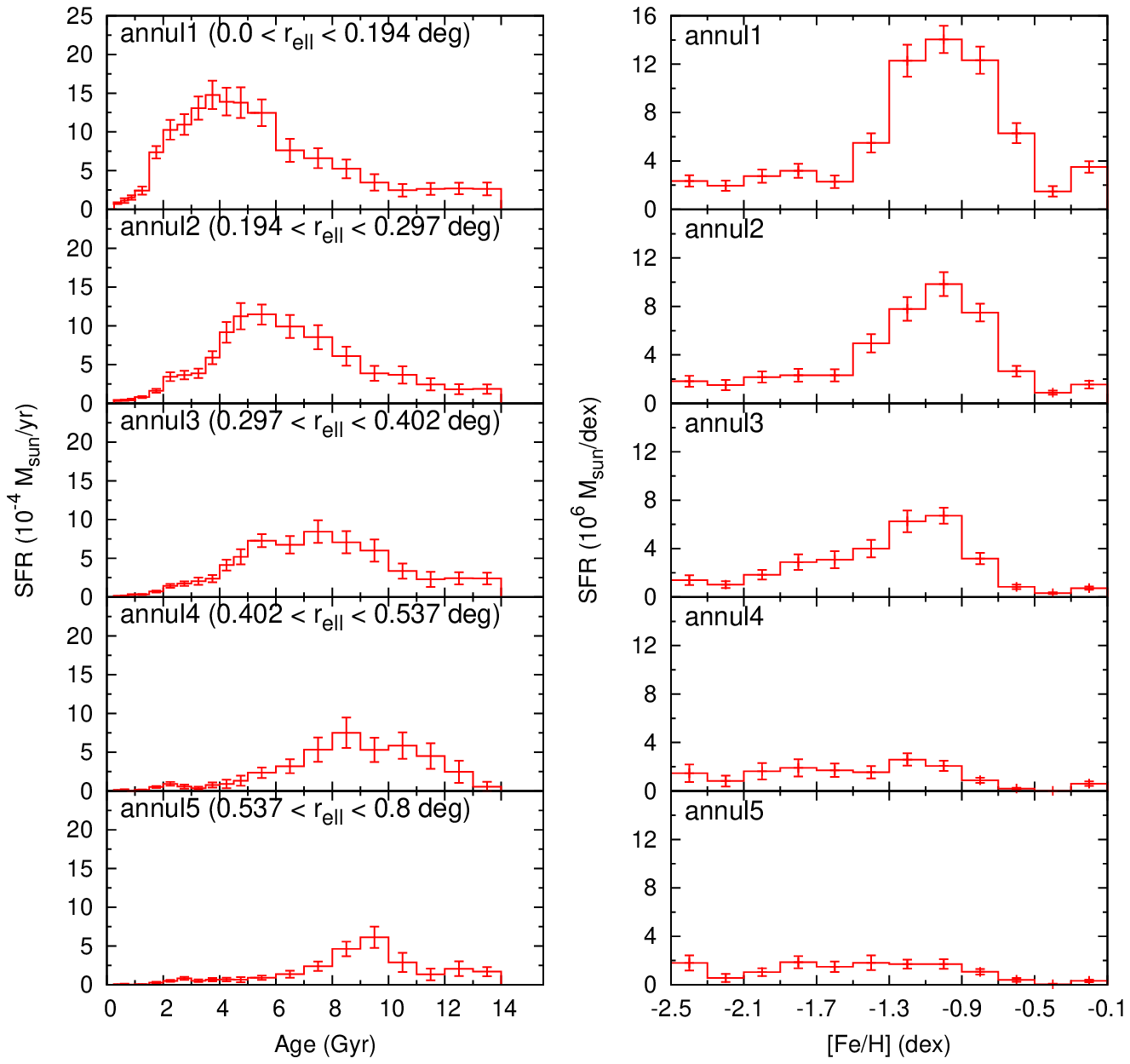}
\caption{The SFH and CEH for 5 annuli of Fornax containing roughly equal number of stars, with the radial extent indicated in each panel. For comparison, the core radius of the Fornax dSph is 0.23 degrees~\citep{Irwin95}. The SFH for annuli 1$-$3 is obtained by fitting all available photometry~(in the B, V and I filters) and spectroscopy.  For annuli 4 and 5 only the V,B-V CMD is used together with the spectroscopy to constrain the SFH. \label{FnxspatialSFH}} 
\end{figure*}

\begin{figure}[!ht]
\centering
\includegraphics[angle=0, width=0.45\textwidth]{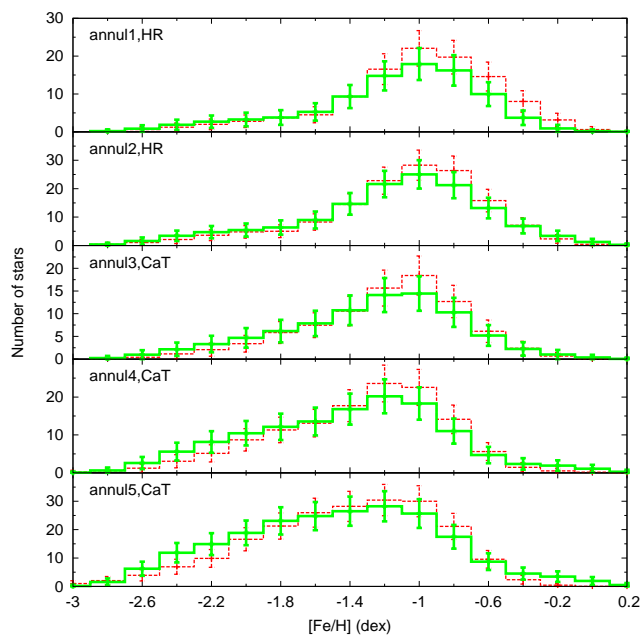}
\caption{Observed~(red, dashed) and  synthetic~(green, solid) histogram MDFs from the SFH analysis in each of the five annuli in Fornax. \label{FnxspatialMDF}} 
\end{figure}

The coverage of the photometric and spectroscopic information extends to a radius of r$_{ell}$=0.8 degrees. Therefore, it is possible to derive the SFH at different radii from the centre, which allows us to determine the presence of variations in the SFH with radius, as seen in Section~\ref{Fnxspatialvariations}. 
\\
The total radial extent of the photometry and spectroscopy is divided into 5 annuli, each containing a similar number of stars observed in the B and V filters, since these filters are fully complete out to greater radii from the centre. For each of the annuli, the SFH and CEH are determined by fitting all available photometric and spectroscopic information, in the same way as discussed in Section~\ref{FnxfullSFH}. For the inner three annuli, photometric information in all three available filters~(B, V and I) is used, together with the spectroscopic MDF. The SFH in the outer two annuli is determined using just the (V, B$-$V) CMD, together with the spectroscopic MDF from~\ion{Ca}{ii} triplet spectroscopy.  \\
Figure~\ref{FnxspatialSFH} shows the best matching SFH and CEH in each of the five annuli of the Fornax dSph. Figure~\ref{FnxspatialSFH}a shows that a clear gradient with age is present in Fornax, with the peak of star formation shifting to progressively older ages for increasing radius. The CEH shown in Figure~\ref{FnxspatialSFH}b shows that the distribution of metallicities changes from peaked in the inner parts to a broad distribution in the outermost bin. \\
The young star formation~(age$\le$1 Gyr) is present only within the centre of Fornax, consistent with the observed CMDs. This is also seen in the CEH, where significant star formation at high metallicities~([Fe/H]$\ge$-0.5 dex) is found only in the central part of Fornax. Very low levels of young star formation are seen in the outer parts, which are likely due to fitting the probable BSS population in the outer parts as a young stellar population. \\
The intermediate age population of Fornax displays a clear gradient in age with radius from the centre. The peak of star formation shifts from an age of $\approx$8-10 Gyr in the outermost annulus to a peak at age$\approx$4 Gyr in the central bin. The CEH of Fornax changes from a broad distribution in the outer regions to a peak at [Fe/H]$\approx$-1.0 dex in the centre, in good agreement with the dominant RGB population seen in Figures~\ref{FnxBVrad} and~\ref{FnxVIrad}. Given the resolution of the SFH solution at these ages~(see Section~\ref{Fnxageresolution}), this radial change in the SFH corresponds to a smoothly varying SFR with radius in Fornax, inconsistent with distinct short bursts of star formation. \\
Old star formation~(age$\ge$10 Gyr) is present at all radii in Fornax, with a flat distribution as a result of the poor SFH resolution. The SFR starts to increase gradually in all regions at an age of $\approx$10 Gyr, consistent with previous studies~\citep{Buonanno99, Gallart052}. \\
The observed and synthetic MDFs inferred from the SFH are shown in Figure~\ref{FnxspatialMDF}, for all five radial annuli. The synthetic MDFs are consistent with the observed spectroscopic MDFs within the errors, as expected, given that the MDF is used as an input in the SFH determination. In the outer parts of Fornax, the metal-poor component is slightly over-estimated in the synthetic MDF, while the inner parts show that the metal-rich populations are slightly under-estimated. The MDF on the upper RGB shows the same trend with radius as the CEH in Figure~\ref{FnxspatialSFH}b, changing from a peaked distribution in the innermost bin to a more broad distribution in annulus 5, consistent with the diminishing strength of the dominant intermediate age component. \\
The total mass in stars formed in our SFH of the Fornax dSph is 4.3$\times$10$^{7}$ M$_{\odot}$ within 0.8 degree or 1.9 kpc. The core radius of the mass distribution resulting from the SFH solution is r$_{c}$=0.29$\pm$0.03 degrees, determined using a Sersic profile fit. This is consistent within the errors with the core radius derived from the observed density profile~\citep{Battaglia06}.

\section{The timescale for chemical evolution of the Fornax dSph}
\label{Fnxindivages}
Using the SFH presented in Figures~\ref{FnxoverallSFH} and~\ref{FnxspatialSFH} it is possible to determine the age probability function of individual stars on the RGB. For each observed star, all stars in the synthetic CMD~(generated using the SFH) with the same magnitude~(in all filters) and metallicity within the observed uncertainties are selected. The ages of these synthetic stars build up a probability distribution function for age, which is considered to be representative of the age of the observed star, as described in~\citet{deBoer2012A}. Assuming a Gaussian distribution, the mean of the distribution function is adopted as the age of the observed star, with the standard deviation used as an error bar. Using this method, we determine accurate age estimates for all available samples of spectroscopic stars in Fornax~\citep{Battaglia082, Starkenburg10, Letarte10, Kirby10}.

\subsection{Age-metallicity relation}
\begin{figure}[!htb]
\centering
\includegraphics[angle=0, width=0.47\textwidth]{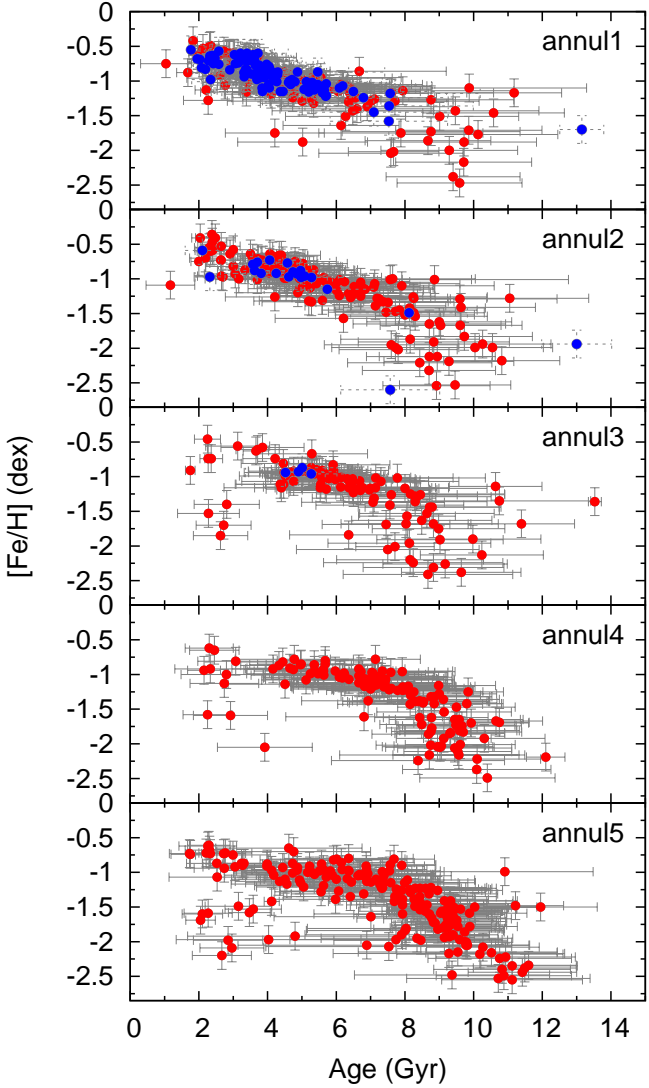}
\caption{The Age-Metallicity Relation of stars on the upper RGB in Fornax, in each of the five annuli, from medium and high resolution spectroscopy~\citep{Letarte10, Kirby10} as blue points, and~\ion{Ca}{ii} triplet spectroscopy~\citep{Battaglia082} as red points. \label{FnxAMRspatial}} 
\end{figure}

Using the accurate ages for stars with spectroscopic metallicities, we determine the detailed AMR of the Fornax dSph~(Figure~\ref{FnxAMR}). The results are consistent with previous determinations of the AMR~\citep{Pont04, Battaglia06}, but determine the age of stars with greater accuracy, revealing the detailed effects of different populations. The AMR shows that the metallicity in Fornax increased rapidly from [Fe/H]$\le$$-$2.5 dex to [Fe/H]=$-$1.5 dex between~8$-$12 Gyr ago. However, the poor SFH resolution results in a large scatter in age, removing any features at the oldest ages. \\
At intermediate ages~(3-8 Gyr), the interstellar medium is gradually enriched during several gigayears, resulting in a narrow, well-defined AMR which reaches [Fe/H]$\approx$$-$0.8 dex at $\approx$3 Gyr. Furthermore, from Figure~\ref{FnxAMR}, hints are seen that the slope of the AMR may become steeper at even younger ages~($\le$3 Gyr), corresponding to a phase of rapid enrichment~\citep{Pont04}. However, more spectroscopic observations, in particular of young BP stars, are needed to unambiguously detect this change in slope. The young, metal-poor stars in the lower left of Figure~\ref{FnxAMR} most likely correspond to AGB stars which are fit as if they were RGB stars~(due to the absence of the AGB phase in the adopted isochrone set), leading to a too young age estimate for these stars. \\ 
A comparison to the AMR of the Sculptor dSph shows that the slope in Fornax at the oldest ages is steeper than in Sculptor, which indicates that metal enrichment occurred with a different rate. After the initial phase of metal enrichment, the slope of the AMR is comparable in both galaxies, but Sculptor only shows stars with old~($>$7 Gyr) ages. \\ 
Figure~\ref{FnxAMRspatial} shows the spatially resolved AMR of Fornax, in each of the five annuli described in Section~\ref{Fnxregions}. The shape of the AMR is consistent in all annuli, with a similar slope at intermediate ages. At the oldest ages, more scatter is seen in the inner two annuli than in the outer three annuli. The presence of the age gradient is visible in Figure~\ref{FnxAMRspatial} as an increased number of stars with younger ages in the AMR for more centrally concentrated regions. 

\subsection{Chemical abundances}
\begin{figure}[!htb]
\centering
\includegraphics[angle=0, width=0.47\textwidth]{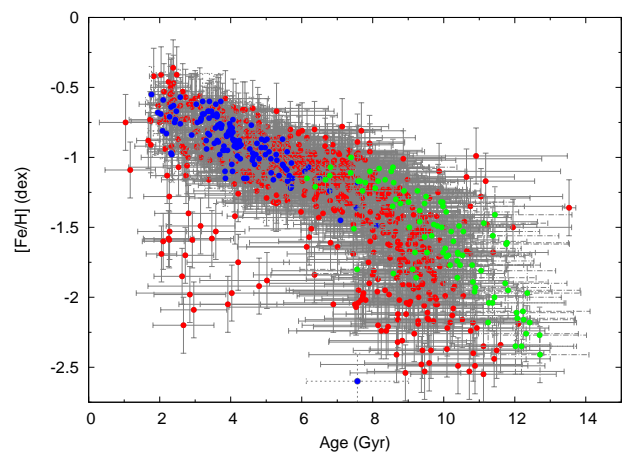}
\caption{The Age-Metallicity Relation of stars on the upper RGB in the Fornax dSph, incorporating the full SFH and MDF information. Medium and high resolution spectroscopy from~\citet{Letarte10, Kirby10} is shown as blue points, while~\ion{Ca}{ii} triplet spectroscopy from~\citet{Battaglia082} is shown in red. The AMR of the Sculptor dSph, from HR spectroscopy, is shown as green points~\citep{deBoer2012A}. \label{FnxAMR}} 
\end{figure}

The accurate age estimates can also be coupled to the stars for which other elements are determined, to derive the evolution of individual elements as a function of time. In this way, we can directly measure the detailed timescale for chemical evolution in Fornax. Figure~\ref{FnxMgFeage}a shows the [Mg/Fe] measurements for individual RGB stars as a function of [Fe/H], with the age of the individual stars colour coded, and Figure~\ref{FnxMgFeage}b shows [Mg/Fe] directly as a function of age. 
\\
Figure~\ref{FnxMgFeage} shows the effect of different stellar populations on [Mg/Fe] in Fornax, and the rate with which this abundance changes at different times. Metal-poor stars are not well sampled in Figure~\ref{FnxMgFeage}, but the most metal-poor stars are consistent with the abundances observed for the MW halo~(Lemasle et al., in prep). Conversely, the metal-rich stars are different from the MW halo, which indicates that Fornax experienced a different evolutionary history than the systems that may have formed the MW halo, as seen in other Local Group dwarf galaxies~\citep{Tolstoy09}. The metal-rich stars in Figure~\ref{FnxMgFeage}a display rapidly decreasing [Mg/Fe] abundances, with a clear trend in [Fe/H] and age. Hints are seen of a changing slope in the [Mg/Fe] distribution for populations of different ages. \\
Connecting the metal-poor and metal-rich populations observed in the [Mg/Fe] distribution in Figure~\ref{FnxAMR}, a ``knee" may be expected to occur at [Fe/H]$\approx$$-$1.5 dex, at an age of 7$-$10 Gyr. The ``knee" in [Mg/Fe] marks the time at which SNe Ia start to contribute to the [Fe/H] content of a galaxy~\citep{Tinsley79, Matteucci90, Gilmore91, Matteucci03}. However, due to the limited sampling of intermediate metallicity stars in Fornax, the ``knee" is not visible in the data presented in Figure~\ref{FnxMgFeage}. Additionally,the large range in [Fe/H] covered by the metal-rich stars in Figure~\ref{FnxMgFeage}a indicates that the ``knee" will likely be spread over a large metallicity range, which may be a result of the extended SFH and multiple populations of Fornax. 

\section{Discussion}
\label{Fnxdiscussion}
We have presented the detailed SFH and CEH of the Fornax dSph out to r$_{ell}$=0.8 degrees, obtained using a combination of deep, multi-colour photometry and spectroscopic metallicities. The obtained, spatially resolved SFH~(see Figures~\ref{FnxoverallSFH} and~\ref{FnxspatialSFH}) shows features similar to previous SFHs~\citep[e.g.][]{Gallart052, Coleman08}. However, the SFH presented here resolves the stellar ages with greater accuracy, due to the inclusion of the spectroscopic MDF. In particular, the age of the dominant episode of star formation is determined more precisely, consistent with the position of the RGB in the observed CMD~(see Section~\ref{FnxRGB}). 
\\
The SFH of Fornax shows little variation in SFR at old ages~($>$10 Gyr), and lacks a dominant episode of star formation at the oldest ages, contrary to what was observed by~\citet{Coleman08}. The absence of a peak in the SFR at the start of star formation is likely an effect of the age resolution of the SFH, smoothing out any features at the oldest ages. To determine if the SFR peaks at the oldest age, deeper photometric data is needed. \\
Fornax displays an extended SFH, dominated by star formation at intermediate age~(1$-$10 Gyr). The reason for the occurrence and strength of this intermediate age episode is not clear. One possibility is a merger with a gas-rich companion, which triggered a strong episode of star formation. However, this gas would need to be pre-enriched at a relatively high metallicity~([Fe/H]$\approx$$-$1.5 dex). Another scenario is that initially expelled gas remained bound to the system for an extended period of time, and fell back onto the galaxy, leading to an episode of star formation similar to what is predicted by simulations~\citep{Revaz09}. The fall-back of expelled gas could be caused by an interaction with the MW~\citep{Piatek07}, similar to models proposed for the Carina dSph~\citep{Pasetto11}. Furthermore, the re-accretion of gas can also be a result of the orbit shape of Fornax, compressing and re-gathering expelled gas at apogalacticon, allowing for repeated bursts of star formation over the course of several orbits~\citep{Nichols12}.
\\
Figure~\ref{FnxoverallSFHdetail} shows that star formation was still taking place in the very centre of Fornax a few 100 Myr ago, resulting in a visible young population~(see Figure~\ref{FnxBVrad}), which ranges in age between $\approx$2 Gyr and 0.25 Gyr. The age of the youngest stars may be younger than can be fit using the adopted isochrone set, which would lead to an incorrect estimate of the SFR in the youngest bin~(0.25$-$0.50 Gyr). However, the level of star formation is expected to be low, not significantly influencing the results at young ages in Figure~\ref{FnxspatialSFH}. \\ 
The young stars of Fornax display a different spatial distribution with respect to the old stars. Furthermore, the ages of these young stars overlap with those of the shell features found in the Fornax dSph~\citep{Coleman04, Coleman05}. Therefore, a link may be present between these features, as suggested by~\citet{Coleman05}. The infall of a smaller system or of gas may be responsible for the substructures, and may have triggered centrally concentrated star formation with a different spatial distribution. To investigate this link, the detailed kinematics and ages of young stars in Fornax need to be determined.

\subsection{Comparison to the Sculptor dSph}
It is useful to compare the results derived for Fornax with our previous study of the Sculptor dSph~\citep{deBoer2012A}. Sculptor is a very different galaxy, which experienced a much shorter evolution history, leading to a different SFH and abundance pattern. However, the same types of measurements were obtained in Fornax and Sculptor, allowing a detailed comparison between the properties of both galaxies. \\
A comparison between the AMR of Fornax and Sculptor~(see Figure~\ref{FnxAMR}) suggests that the initial metal enrichment in Fornax occurred on a shorter timescale than in Sculptor, leading to a steeper slope at earlier times. A possible explanation for this could be the greater mass of Fornax, which allows it to retain more of its gas after SNe explosions, leading to more rapid enrichment, similar to what is predicted by simulations~\citep[e.g., ][]{Revaz11}. This rapid metal enrichment could also be responsible for the different position of the [Mg/Fe] ``knee" in Fornax, compared to the observed ``knee" in Sculptor~\citep{deBoer2012A}. \\
The dominant stellar component of the Fornax dSph is an intermediate age population. This is very different to Sculptor, which formed the bulk of its stars before $\approx$10 Gyr. This lack of intermediate age star formation may be due to its lower~(dynamical) mass, which did not allow it to retain as much of its gas as Fornax and prevented it from continuing to form stars. 
\begin{figure*}[!htb]
\centering
\includegraphics[angle=0, width=0.49\textwidth]{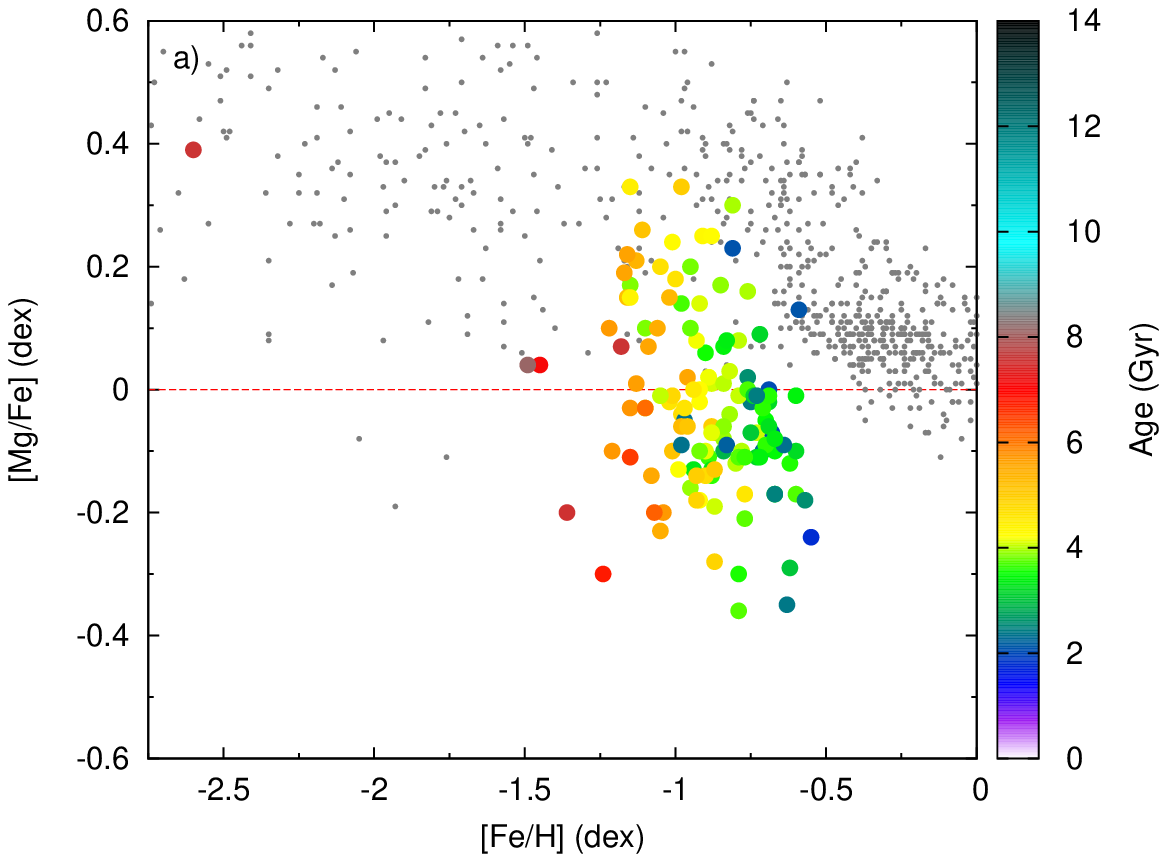}
\includegraphics[angle=0, width=0.49\textwidth]{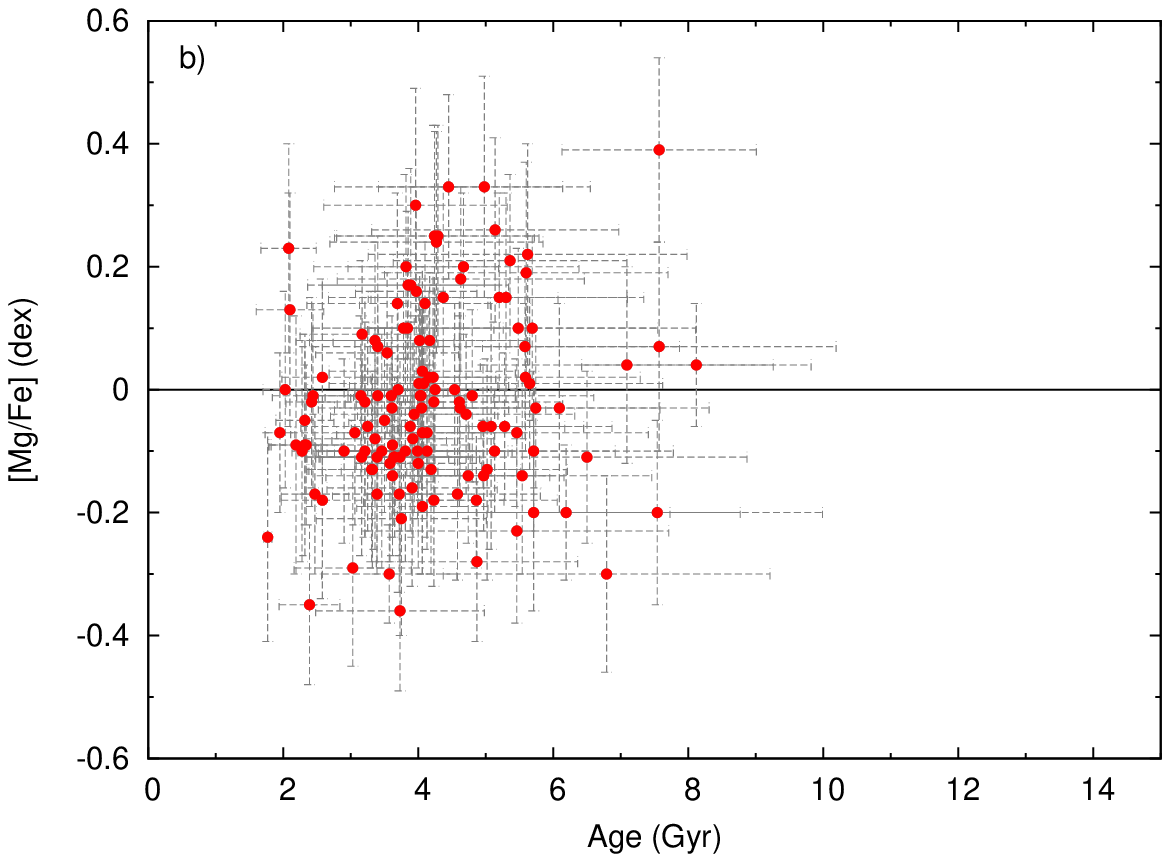}
\caption{\textbf{a)} [Mg/Fe] measurements for the medium and high resolution spectroscopic samples of RGB stars from~\citep{Letarte10, Kirby10}~(coloured filled circles). The colours represent the age in Gyr, derived from the SFH. Stars in the Milky Way are shown for comparison~(small grey points). \textbf{b)} [Mg/Fe] is plotted directly against age for the same sample. \label{FnxMgFeage}} 
\end{figure*}

\section{Summary}
\label{FnxSummary}
We have presented deep B, V and I CMDs of the Fornax dSph, out to an elliptical radius of 0.8 degrees, containing a total of 270.000 stars. The CMDs show that Fornax displays a radial gradient with age and metallicity, with more metal-rich, younger stars forming more toward the centre. Fornax is dominated by intermediate age~(1$-$10 Gyr) star formation, and shows a strong dominant RGB linked to a stellar population with an age of $\approx$4 Gyr, at [Fe/H]$\approx$$-$1.0 dex. \\
Combining the deep CMDs with the spectroscopic MDF of individual RGB stars we have derived the detailed SFH of Fornax, at different radii from the centre. The spatially resolved SFH confirms and quantifies the radial age and metallicity gradient, with a peak of star formation that shifts to older ages for increasing radius. From the SFH we determine that the total mass in stars formed in Fornax is 4.3$\times$10$^{7}$ M$_{\odot}$ within an elliptical radius of 0.8 degrees, or 1.9 kpc. \\
Using the SFH, we determined age estimates for individual RGB stars, for which spectroscopic abundances are available. This allows us to study, for the first time, the evolution of different elements in Fornax, directly as a function of time. Using the individual age estimates, we determined the detailed AMR of individual stars and timescale for evolution of $\alpha$-elements of the Fornax dSph. The AMR shows that Fornax experienced an initial period of star formation, which enriched the metallicity from [Fe/H]$\le$$-$2.5 dex to [Fe/H]=$-$1.5 dex from $\approx$12 to 8 Gyr ago. Subsequent gradual metal enrichment resulted in a narrow, well-defined AMR during the next 5 Gyr that reaches [Fe/H]$\approx$$-$0.8 dex at $\approx$3 Gyr. Hints are seen in the AMR of a second change in slope at young ages~(1$-$3 Gyr). \\
The detailed timescale of chemical enrichment~(see Figure~\ref{FnxMgFeage}) shows that the abundances of old, metal-poor stars are consistent with the MW, while the abundances of more metal-rich stars show a departure from MW abundances. Metal-rich~([Fe/H]$\approx$$-$1.3 dex) stars show a rapid decrease of [Mg/Fe], with a clear trend with age. Connecting both features predicts the presence of a knee in the alpha-element distribution at [Fe/H]$\approx$$-$1.5 dex, at 7$-$10 Gyr. Hints are seen of a different slope in the [Mg/Fe] distribution for different populations, which could be a result of the different strength of previous stellar populations, leading to a different rate of chemical enrichment. 

\section{Acknowledgements}
\label{acknowledgements}
The authors thank ISSI (Bern) for support of the team ``Defining the full life-cycle of dwarf galaxy evolution: the Local Universe as a template". T.d.B and E.T. gratefully acknowledge the Netherlands Foundation for Scientific Research (NWO) for financial support through a VICI grant. E.O. is partially supported by NSF grant AST0807498. Partial support was provided to E.S. by a CITA National Fellowship and CIfAR's Junior Academy. The authors would like to thank the anonymous referee for his/her comments, that helped to improve the paper.

\bibliographystyle{aa}
\bibliography{Bibliography}

\clearpage
\begin{appendix}
\onecolumn
\section{Observing log}
\label{Fnxobslist}
\begin{table*}[!h]
\caption[]{List of observed fields in the Fornax dSph, with the 4m CTIO Blanco telescope during two observing runs. Information is given about exposure time, airmass and seeing conditions as determined on image. \label{Fnx4mstds}}
\footnotesize
\begin{center}
\begin{tabular}{lccccccc}
\hline\hline
Date 	& Field		&  RA		& DEC		& Filter	& exp time	& seeing	& airmass  \\
 	&				& J2000		& J2000		& 		& sec		& $^{\prime\prime}$ &        \\	
\hline
2009 Nov 19	   & Fnxc1 & 02:39:51.84   & $-$34:47:48.96 & B & 3600, 90, 10   & 1.2        & 1.08-1.25	\\
2009 Nov 19 	   &             &                          &                              & V & 3600, 90, 10   & 1.2-1.4 & 1.30-1.70	\\
2009 Nov 22 	   &             &                          &                              & I  & 4800, 90, 10   & 1.0-1.5 & 1.25-1.70	\\
2009 Nov 21 	   & Fnxc2 & 02:39:51.84   & $-$34:13:48.96  & B & 3600, 90, 10   & 0.9-1.2 & 1.08-1.26	\\
2009 Nov 21 	   &             &                          &                              & V & 3600, 90, 10   & 1.1-1.5 & 1.30-1.70	\\
2009 Nov 23 	   &             &                          &                              & I  & 4800, 90, 10   & 1.0-1.3 & 1.20-1.68	\\
2008 Oct 4	   & FnxNE & 02:42:07.28  & $-$34:03:34.90  & B & 6600, 90, 10   & 0.9-1.7 & 1.07-1.55	\\
2008 Oct 4 	   &             &                          &                              & V & 9000, 90, 10   & 1.0-1.5 & 1.09-1.74	\\
2008 Oct 6 	   & FnxNW & 02:37:25.24 & $-$34:11:44.40  & B & 4200, 90, 10  & 1.3-1.6 & 1.00-1.21	\\
2008 Oct 6 	   &             &                          &                              & V & 7800, 90, 10   & 1.3-1.6 & 1.00-1.15	\\
2008 Oct 5-6	   & FnxSE & 02:42:28.77   & $-$34:42:00.10  & B & 5400, 90, 10  & 1.1-1.7 & 1.00-1.70	\\
2008 Oct 5-6 	   &             &                          &                              & V & 7200, 90, 10   & 0.9-1.6 & 1.00-1.80	\\
2008 Oct 4-5 	   & FnxSW & 02:37:52.57 & $-$34:50:12.30  & B & 4200, 90, 10   & 1.0-1.4 & 1.08-1.43	\\
2008 Oct 4-5 	   &             &                          &                              & V & 7200, 90, 10   & 0.9-1.3 & 1.06-1.60	\\
2008 Oct 5-6	   & FnxIW & 02:39:51.84   & $-$34:47:48.96  & B & 1800    & 1.2-1.4 & 1.09-1.20	\\
2008 Oct 5-6 	   &             &                          &                              & V & 6000     & 1.3-1.6 & 1.06-1.30	\\
\hline 
\end{tabular}
\end{center}
\end{table*}

\end{appendix}

\end{document}